\documentclass[prx,twocolumn,floatfix,superscriptaddress,amsmath,aps,longbibliography]{revtex4-2}
\pdfoutput=1
\usepackage{dcolumn,graphicx,color,booktabs,microtype,afterpage} \graphicspath{{figures/}}
\makeatletter\renewcommand{\fnum@figure}[1]{\figurename~\thefigure.}\makeatother
\makeatletter\renewcommand{\fnum@table}[1]{\tablename~\thetable.}\makeatother

\newcommand{\TN}{$T_{\mathrm{N}}$}
\newcommand{\TSR}{$T_{\mathrm{SR}}$}

\usepackage[colorlinks,plainpages=false,linkcolor=blue,urlcolor=blue,citecolor=blue,pdfpagemode=UseNone,pdfstartview=FitBH]{hyperref}

\begin{document}

\title{Controlling the Temperature of the Spin\textendash Reorientation Transition 
In~HoFe$_{1 \textendash x}$Mn$_x$O$_3$ Orthoferrite Single Crystals}
% Force line breaks with \\

\author{K. A. Shaykhutdinov}
\thanks{Corresponding author: smp@iph.krasn.ru}
\affiliation{Kirensky Institute of Physics, Federal Research Center KSC SB RAS, Krasnoyarsk, 660036 Russia}
\author{S. A. Skorobogatov}
\affiliation{Kirensky Institute of Physics, Federal Research Center KSC SB RAS, Krasnoyarsk, 660036 Russia}
\author{Yu. V. Knyazev}
\affiliation{Kirensky Institute of Physics, Federal Research Center KSC SB RAS, Krasnoyarsk, 660036 Russia}
\author{T. N. Kamkova}
\affiliation{Kirensky Institute of Physics, Federal Research Center KSC SB RAS, Krasnoyarsk, 660036 Russia}
\affiliation{Department of Solid State Physics and Nanotechnology, Institute of Engineering Physics and Radioelectronics, Siberian Federal University, Krasnoyarsk 660041, Russia}
\author{A. D. Vasil’ev}
\affiliation{Kirensky Institute of Physics, Federal Research Center KSC SB RAS, Krasnoyarsk, 660036 Russia}
\affiliation{Department of Solid State Physics and Nanotechnology, Institute of Engineering Physics and Radioelectronics, Siberian Federal University, Krasnoyarsk 660041, Russia}
\author{S. V. Semenov}
\affiliation{Kirensky Institute of Physics, Federal Research Center KSC SB RAS, Krasnoyarsk, 660036 Russia}
\affiliation{Department of Solid State Physics and Nanotechnology, Institute of Engineering Physics and Radioelectronics, Siberian Federal University, Krasnoyarsk 660041, Russia}
\author{M. S. Pavlovskii}
\affiliation{Kirensky Institute of Physics, Federal Research Center KSC SB RAS, Krasnoyarsk, 660036 Russia}
\affiliation{Department of Solid State Physics and Nanotechnology, Institute of Engineering Physics and Radioelectronics, Siberian Federal University, Krasnoyarsk 660041, Russia}
\author{A. A. Krasikov}
\affiliation{Kirensky Institute of Physics, Federal Research Center KSC SB RAS, Krasnoyarsk, 660036 Russia}

\begin{abstract}

HoFe$_{1 \textendash x}$Mn$_x$O$_3$ $(0 < x < 1)$ single crystals have been grown by the optical floating zone technique. A structural transition from the orthorhombic to hexagonal modification has been established in the crystals in the concentration range of $0.7\textendash 0.8$, which has been confirmed by the X\textendash ray diffraction data. For a series of the rhombic crystals, the room-temperature Mössbauer study and magnetic measurements in the temperature range of $4.2\textendash 1000$~K have been carried out. It has been observed that, with an increase in the manganese content in the samples, the temperature of the spin-reorientation transition increases significantly: from 60~K in the HoFeO$_3$ compound to room temperature in HoFe$_{0.6}$Mn$_{0.4}$O$_3$. The magnetic measurements have shown that, upon substitution of manganese for iron, the magnetic orientational type transition changes from a second-order transition ($A_xF_yG_z \rightarrow C_xG_yF_z$) to first-order one ($A_xF_yG_z \rightarrow G_xC_yA_z$) with a weak ferromagnetic moment only in the \textbf{b} direction (for Pnma notation). The growth of the spin-reorientation transition temperature has been attributed to the change in the value of the indirect exchange in the iron subsystem under the action of manganese, which has been found when studying the Mössbauer effect in the HoFe$_{1 \textendash x}$Mn$_x$O$_3$ $(x < 0.4)$ compound.

\end{abstract}

\maketitle{}

\section{Introduction}

Oxide materials containing transition and rare-earth ions exhibit many intriguing effects caused by the complex interplay of two magnetic subsystems. Orthoferrites with the general formula \textit{R}FeO$_3$~\cite{bib1,bib2,bib3,bib4} that can be distinguished into a separate class of such materials have been explored for more than half a century. The \textit{R}FeO$_3$ crystal structure is described by the sp. gr. Pnma (\#62). Although the magnetic properties of these compounds have been thoroughly investigated, interest in them has grown again due to the recent discovery of a number of attracting phenomena: multiferroism below the temperature of ordering of the rare-earth subsystem~\cite{bib5}, laser-induced ultrafast magnetization switching in domain walls~\cite{bib6,bib7,bib8}, formation of a soliton lattice in the TbFeO$_3$ compound \cite{bib9}, and existence of quasi-one-dimensional Yb$^{3+}$ spin chains in the YbFeO$_3$ compound~\cite{bib10}.

Rare-earth orthoferrites represent a family with the extraordinary magnetic phenomena. The unique magnetic properties of these materials result from the complex interplay of the moments of 3d and 4f electrons. It is known well~\cite{bib1} that the \textit{R}FeO$_3$ compounds are characterized by unusually high Néel temperatures (\TN$\sim600\textendash 700$~K), below which the Fe moments are ordered antiferromagnetically with a weak sublattice canting, which induces weak ferromagnetism. As the temperature decreases, the role of the Fe\textendash R interaction increases; as a result, orientational spin transitions \TSR~occur at lower temperatures, which depend strongly on the rare-earth ion: \TSR$\approx 50\textendash 60$~K for HoFeO$_3$~\cite{bib11,bib12}, \TSR$ \approx 80\textendash 90$~K for TmFeO$_3$~\cite{bib11,bib13}, and \TSR$ \approx 3\textendash 10$~K when Tb is used as a rare-earth element~\cite{bib9,bib14}. Of special interest is the SmFeO$_3$ compound ~\cite{bib15}, in which the orientational transition is observed at \TSR $\approx 450\textendash 480$~K. The subsystem of rare-earth ions with a relatively weak R - R interaction at high temperatures is paramagnetic or weakly polarized by the molecular field of ordered Fe ions. The rare-earth magnetic sublattice becomes ordered below \TSR $< 10$~K (the temperature of ordering of the rare-earth subsystem). The observed magnetic properties of the system result from the multiplicity of different exchanges. In addition to the Fe\textendash Fe, Fe\textendash R, and R\textendash R Heisenberg exchange couplings, an important role in determining the magnetic properties is played by the Dzyaloshinskii‒Moriya interaction~\cite{bib16,bib17}, which induces a weak ferromagnetic moment.

A way of controlling the temperature of the spin-reorientation transition is isovalent substitution in the \textit{R}Fe$_{1\textendash x}$\textit{M}$_x$O$_3$ (M = Cr, Mn, Co, Ni) iron subsystem. In this case, it becomes possible to smoothly change the magnetic properties in such systems and set the desired temperature \TSR, for example, at the ultrafast magnetization switching in domain walls~\cite{bib6,bib7,bib8}. For example, in~\cite{bib18}, the reversible spin-reorientation transition was observed in the TbFe$_{0.75}$Mn$_{0.25}$O$_3$ single crystal at \TSR $= 250$~K, whereas in the pure TbFeO$_3$ single crystal, the transition occurs at temperatures of 3 and 8~K. In~\cite{bib19}, a SmFe$_{0.75}$Mn$_{0.25}$O$_3$ single crystal was successfully grown; it was shown that, when a part of iron is replaced by manganese, the temperature \TSR~decreases noticeably. In~\cite{bib20}, the authors managed to grow a series of PrFe$_{1\textendash x}$Mn$_x$O$_3$ ($0<x<0.3$) single crystals and demonstrated that, at a manganese content of 30\%, the spin-reorientation transition shifts to room temperature. In~\cite{bib21}, the synthesis of a series of GdFe$_{1\textendash x}$Mn$_x$O$_3$ ($0<x<0.3$) single crystals was reported. 

It should be noted that the majority of studies have been carried out on polycrystalline samples. Meanwhile, in terms of the potential of the discussed orthoferrites for microelectronic applications, it is desirable to explore their single-crystal samples. We can distinguish studies~\cite{bib22,bib23,bib24} aimed at the examination of dysprosium, terbium, and yttrium orthoferrites with partial substitution of manganese for iron; in these materials, a change in the temperature \TSR~was also observed. In addition, in some works~\cite{bib25,bib26,bib27}, substitution of other transition elements (chromium, nickel, and cobalt) into the iron subsystem was performed and a change in the \TSR value was reported.

Concerning the HoFe$_{1\textendash x}$Mn$_x$O$_3$ compound, in~\cite{bib28} the polycrystalline HoFe$_{1\textendash x}$Mn$_x$O$_3$ ($0<x<0.5$) samples were synthesized and their magnetic and structural features were reported. It was shown that, at a concentration of $x=0.4$, the temperature \TSR~for this compound reached room temperature. However, against the background of holmium paramagnetism, the spin-reorientation transition turns into just a minor anomaly in the magnetization curve. Therefore, to study the evolution of the magnetic properties in more detail, it is necessary to perform measurements on single-crystal samples. 

The aim of this study was to grow and investigate single-crystal HoFe$_{1\textendash x}$Mn$_x$O$_3$ samples with a manganese content of up to its maximum value at which this solid solution remains stable in the orthorhombic phase.

\section{Experimental details}

At the first stage, to obtain the HoFe$_{1\textendash x}$Mn$_x$O$_3$ ($x = 0, 0.05, 0.1, 0.2, 0.3, 0.4, 0.5, 0.6, 0.7,$~and~$0.8$) samples, the initial Ho$_2$O$_3$, Fe$_2$O$_3$, and MnO$_2$ powders (99.9\%, Alfa Aesar) were mixed in required proportions and subjected to annealing at a temperature of 925~°C for 18~h. The annealed powders were poured into a rubber mold and pressed in a hydrostatic press at a pressure of $\sim100$~MPa. The resulting cylindrical samples were then annealed in a vertical furnace at 1400~°C for 16 h. After that, the synthesized polycrystalline HoFe$_{1\textendash x}$Mn$_x$O$_3$ samples were placed in an FZ\textendash T\textendash 4000\textendash H\textendash VIII\textendash VPO\textendash PC optical floating zone furnace (Crystal Systems Corp.) to grow single crystals. The crystal growth occurred in air at a normal pressure and a relative rod rotation speed of 30~rpm. The growth rates varied from 3 to 1~mm/h, depending on the ratio between iron and manganese in the HoFe$_{1\textendash x}$Mn$_x$O$_3$ composition. It should be noted that we attempted to grow the HoFe$_{1\textendash x}$Co$_x$O$_3$, HoFe$_{1\textendash x}$Cr$_x$O$_3$, and HoFe$_{1\textendash x}$Ni$_x$O$_3$ single crystals with $x = 0.05$; however, substitution of even 5\% of chromium or nickel for iron led to incongruent melting and did not allow us to obtain high-quality single crystals. The change in pressure (up to 10~atm) and a gaseous medium (O$_2$, Ar–O$_2$, or Ar–H$_2$) did not improve the growth conditions in this case. It also seemed that the HoFe$_{0.95}$Co$_{0.05}$O$_3$ single crystal was successfully grown, but the measured temperature dependences of the magnetization were found to be the same as for the HoFeO$_3$ crystal, which showed that cobalt was not substituted for iron during the growth.

To solve the crystal structure and determine the phase purity of the HoFe$_{1\textendash x}$Mn$_x$O$_3$ compound, a Bruker SMART APEX II single-crystal X\textendash ray diffractometer was used. The measurements were performed at room temperature. The crystallographic orientations of all the single crystals were determined on a Photonic Science Laue Crystal Orientation System by the back reflection method. Mössbauer spectra for the investigated samples were obtained on an MS\textendash 1104Em spectrometer in the transmission geometry with a $^{57}$Co(Rh) radioactive source at a temperature of 300~K. Temperature and field dependences of the magnetization of the grown single crystals were obtained on a Quantum Design Physical Property Measurement System (PPMS\textendash 6000) in the temperature range of 4.2\textendash 350~K. The high-temperature (350\textendash 1000~K) magnetization measurements were performed on a Lake Shore Cryotronics VSM\textendash 8604 vibrating sample magnetometer. After temperature cycling up to 1000~K, the repeatability of the characteristic (Néel and spin-reorientation) temperatures and the magnetization values were checked. All the measurements were found to be fully repeatable.

\section{Results and analysis}
\subsection{Structure characterization}

\begin{figure}[tb!]
	\centering\includegraphics[width=1\linewidth]{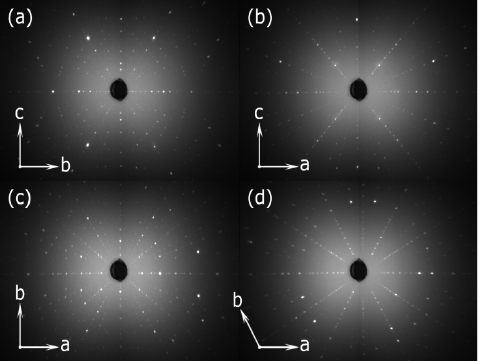}	
	\caption{~Lauegrams of (a, b, c) the HoFe$_{0.7}$Mn$_{0.3}$O$_3$ sample from reflection planes (100), (010), (001) and (d) the HoFe$_{0.2}$Mn$_{0.8}$O$_3$ sample from reflection plane (0001). The direction of the crystallographic axes relative to the reflection plane is shown in each figure.}
 \label{fig1} 
\end{figure}

To check the quality of all the grown HoFe$_{1\textendash x}$Mn$_x$O$_3$ single crystals and their orientations along the three crystallographic axes, the Laue method was used. As an example, Fig.~\ref{fig1} presents Lauegrams of the sample with $x = 0.3$ corresponds to the reflection planes with indices (100), (010), (001) (Figs.~\ref{fig1} (a, b, c)) and the sample with $x = 0.8$ (Fig.~\ref{fig1}d) for the reflection plane (0001). The observed sharp symmetric peaks and the absence of twinning are indicative of the high quality of the synthesized crystals. In addition, the reflections for the sample with $x = 0.3$ correspond to the sp. gr. Pnma (\#62), while the Lauegram of the sample with $x = 0.8$ corresponds to the sp. gr. P63cm (\#185). Thus, upon substitution of manganese for iron in the HoFe$_{1\textendash x}$Mn$_x$O$_3$ ($0.7<x<0.8$) sample, the interface between the orthorhombic and hexagonal modifications in the HoFe$_{1‒x}$Mn$_x$O$_3$ solid solution is observed. We note that the single crystal with $x = 1$ (pure HoMnO$_3$) in the orthorhombic modification can be obtained by the solution‒melt method~\cite{bib29}, which makes it possible to significantly lower the melt temperature, whereas the optical floating zone technique yields only the hexagonal HoMnO$_3$ modification~\cite{bib30}. Based on the obtained data on the interface in HoFe$_{1\textendash x}$Mn$_x$O$_3$, below we present the results of investigations of the orthorhombic HoFe$_{1\textendash x}$Mn$_x$O$_3$($x<0.7$) single crystals synthesized by zone melting under the same conditions.

\begin{figure}[tb!]
	\centering\includegraphics[width=1\linewidth]{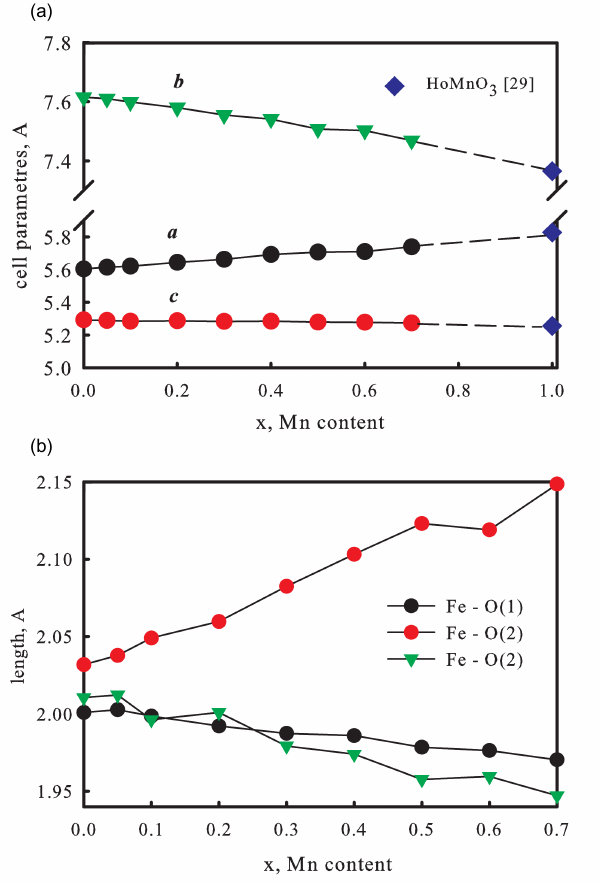}	
	\caption{~(a) HoFe$_{1\textendash x}$Mn$_x$O$_3$ lattice parameters vs manganese content. Blue diamonds show the lattice parameters of the HoMnO$_3$ crystal from~\cite{bib29}. (b) Manganese content dependence of Fe\textendash O(1) and Fe\textendash O(2) bond lengths in the iron‒oxygen octahedron.}
 \label{fig2} 
\end{figure}

It was established from the single-crystal X\textendash ray diffraction data that the HoFe$_{1\textendash x}$Mn$_x$O$_3$ ($x<0.7$) single crystals are orthorhombic with the sp. gr. Pnma. Their lattice parameters are given in Table~\ref{table1} and presented in Fig.~\ref{fig2}.

\begin{table}[tb!]
\caption {~Lattice parameters, unit cell volumes in the HoFe$_{1\textendash x}$Mn$_x$O$_3$ compound.}
\label{table1}
\begin{ruledtabular}
\begin{tabular}{ll@{\qquad} @{\qquad}llll}
& $x$ & $a$ & $b$ & $c$ & $V$\\ 
\hline
 & $0$ & $5.6029(5)$ & $7.6151(7)$ & $5.2921(5)$ & $225.80(5)$\\
 & $0.05$ & $5.6133(3)$ & $7.6102(5)$ & $5.2885(3)$ & $225.92(2)$\\
 & $0.1$ & $5.6201(5)$ & $7.5990(4)$ & $5.2840(3)$ & $225.66(2)$\\
 & $0.2$ & $5.6425(3)$ & $7.5801(4)$ & $5.2864(3)$ & $226.10(2)$\\
 & $0.3$ & $5.6608(3)$ & $7.5551(3)$ & $5.2826(2)$ & $225.93(2)$\\
 & $0.4$ & $5.6911(4)$ & $7.5411(5)$ & $5.2840(4)$ & $226.77(3)$\\
 & $0.5$ & $5.7065(3)$ & $7.5080(4)$ & $5.2779(3)$ & $226.73(3)$\\
 & $0.6$ & $5.7085(3)$ & $7.5029(3)$ & $5.2768(2)$ & $226.01(2)$\\
 & $0.7$ & $5.7385(3)$ & $7.4688(4)$ & $5.2728(3)$ & $225.99(3)$\\ 
\end{tabular}
\end{ruledtabular}
\end{table}

\begin{figure}[tb!]
	\centering\includegraphics[width=1\linewidth]{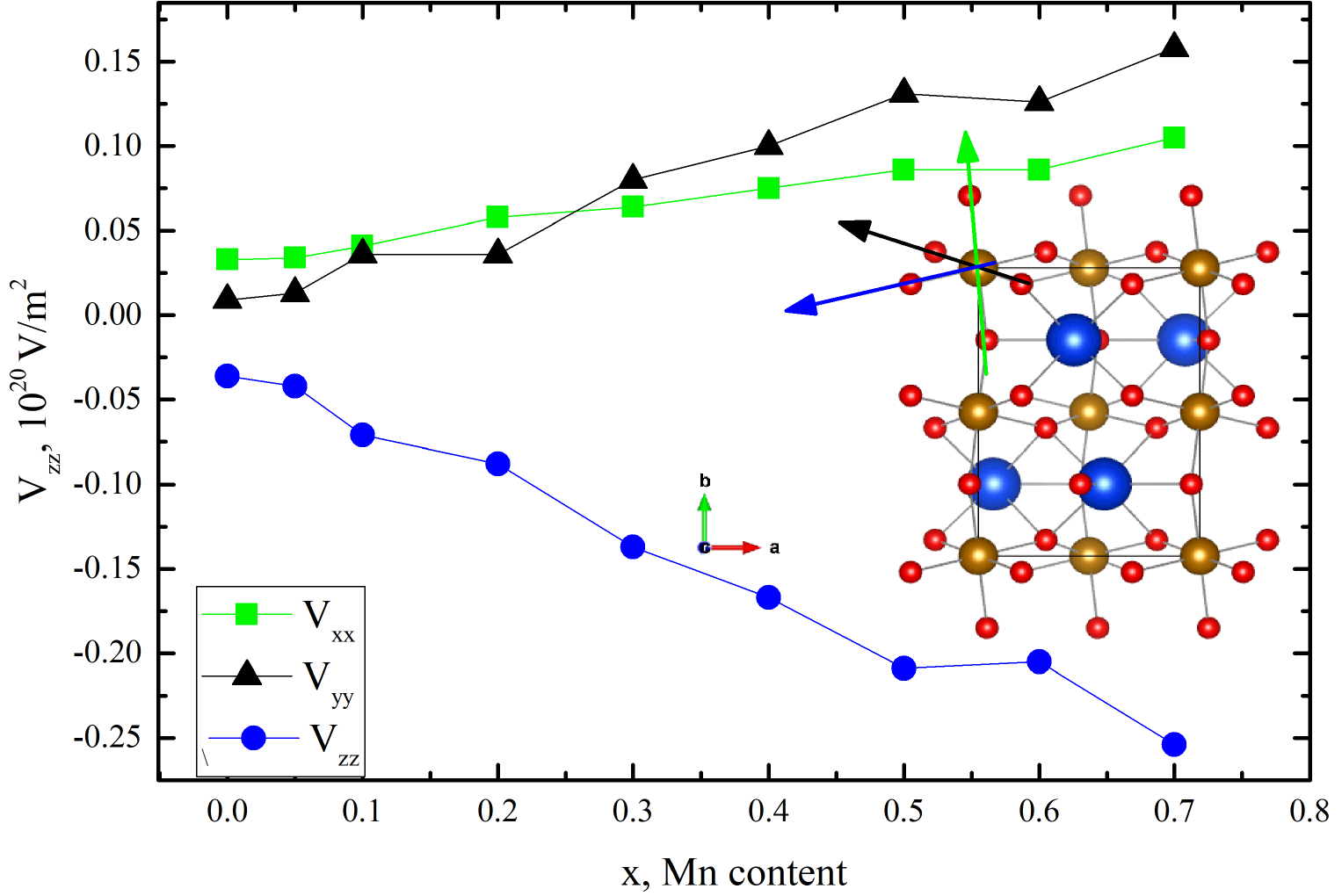}	
	\caption{~EFG tensor components vs Mn content in the HoFe$_{1\textendash x}$Mn$_x$O$_3$ compound. The inset shows the directions of all the components of the EFG vector relative to the crystallographic axes.}
 \label{fig3} 
\end{figure}

It can be seen from Table~\ref{table1} and Fig.~\ref{fig2} that, under the isovalent substitution of manganese for iron in the HoFe$_{1‒x}$Mn$_x$O$_3$ ($x<0.7$) compound, the lattice parameters change linearly over the entire concentration range. In this case, parameter a increases linearly, parameter \textbf{b} decreases, and parameter c decreases insignificantly; the unit cell volume remains almost unchanged. Figure~\ref{fig3} presents the concentration dependence of Fe\textendash O(1) and Fe\textendash O(2) bond lengths in the iron\textendash oxygen octahedron. It can be seen that, as the $x$ value increases, the octahedron shrinks along the b direction and the greatest changes occur in the octahedron plane close to the ac plane, where the Fe\textendash O(2) bond significantly elongates. Using the single\textendash crystal X\textendash ray diffraction data, we can determine the direction of the main component of the electric field gradient (EFG) tensor $V_{zz}$, which conventionally determines the principal axis direction in the octahedron. The $V_{zz}$  can be determined from the X\textendash ray diffraction data using the known bond lengths and angles in the octahedral environment of iron in the nearest neighbor approximation~\cite{bib31} as

\begin{eqnarray}
   V_{zz}=\sum 2e \frac{2cos^2(\theta)-1}{r^3}
    \label{Eq:eq1}
\end{eqnarray}

Where $V_{zz}$ is the EFG in the direction of the principal axis of the oxygen octahedron, $\theta$ is the angle between this axis and the direction to the neighboring oxygen ion, $e$ is the elementary charge, and $r$ is the metal\textendash oxygen distance. In the calculation, all possible directions of the EFG axis in the local environment of the cation are checked, after which the main component of the EFG tensor is selected according to the condition$|{V_{zz}}| \geq |V_{yy}| \geq |V_{xx}|$.

\begin{table}[tb!]
\caption {~Metal\textendash oxygen distances at different Mn contents.}
\label{table2}
\begin{ruledtabular}
\begin{tabular}{ll @{\qquad} @{\qquad} lll}
& $x$ & $M-O1(4c)$ & $M-O2(8d)$ & $M-O2(8d)$\\ 
\hline
 & $0$ & $2.001$ & $2.011$ & $2.032$\\
 & $0.05$ & $2.003$ & $2.012$ & $2.038$\\
 & $0.1$ & $1.998$ & $1.996$ & $2.049$\\
 & $0.2$ & $1.992$ & $2.001$ & $2.060$\\
 & $0.3$ & $1.987$ & $1.979$ & $2.082$\\
 & $0.4$ & $1.986$ & $1.974$ & $2.103$\\
\end{tabular}
\end{ruledtabular}
\end{table}

\begin{figure}[tb!]
	\centering\includegraphics[width=1\linewidth]{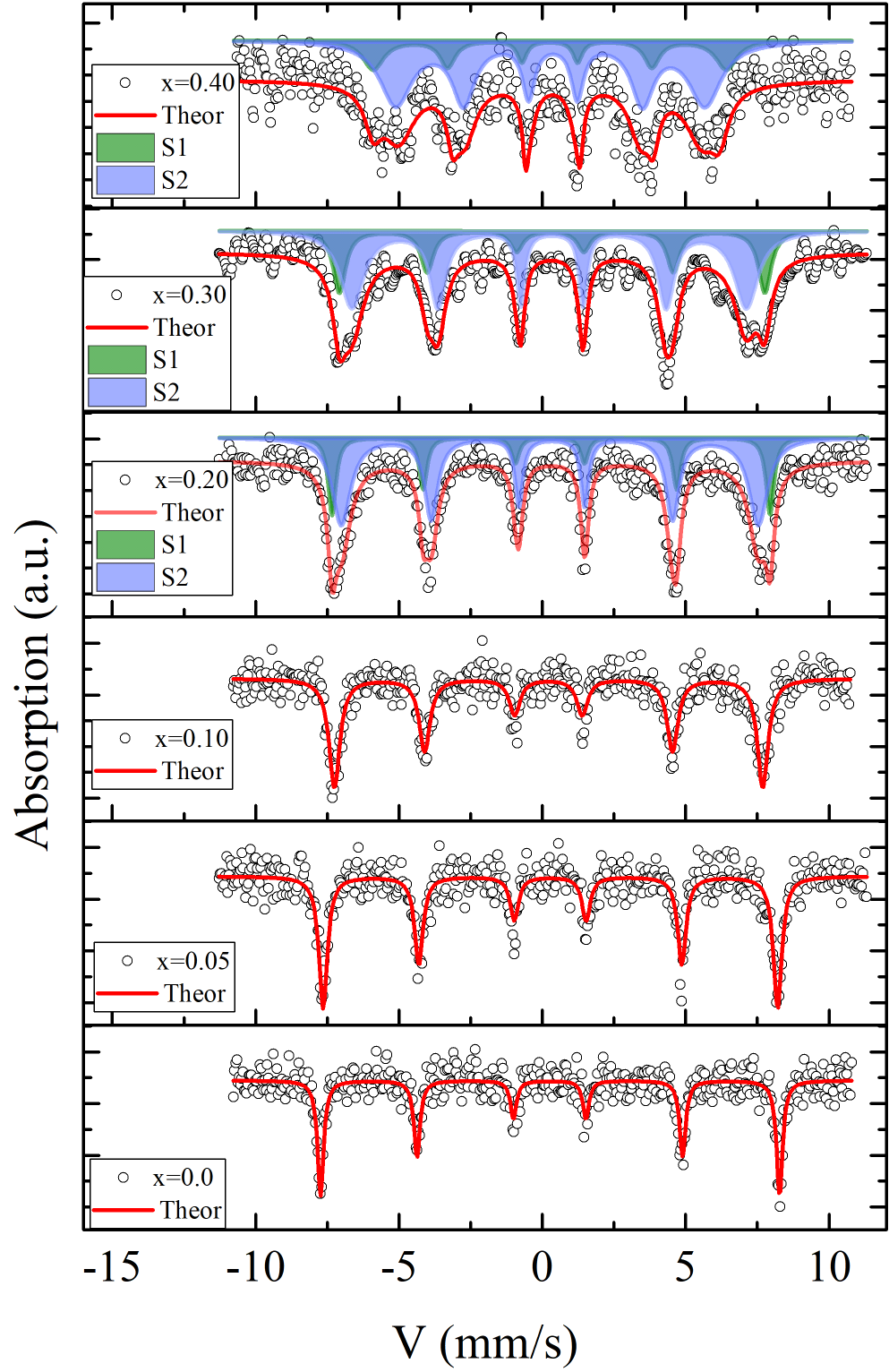}	
	\caption{~Mössbauer spectra of the HoFe$_{1\textendash x}$Mn$_x$O$_3$ ($x=0, 0.05, 0.10, 0.20, 0.30,$~and~$0.40$) samples. The solid line shows the processing results. The shaded areas show partial components in the spectra.}
 \label{fig4} 
\end{figure}

The results of the calculation in all possible directions for all the samples are given in Table~\ref{table2} and presented in Fig.~\ref{fig4}. Thus, we determine the direction of the main component of the EFG tensor relative to the crystallographic axes (shown in blue in the inset to Fig.\ref{fig4}). The chemical bond length in this direction exceeds the lengths in the two other directions, so the oxygen octahedron is elongated.
 
\subsection{Mössbauer Spectroscopy Study}

Samples for the Mössbauer spectroscopy study were prepared by grinding the HoFe$_{1\textendash x}$Mn$_x$O$_3$ ($0<x<0.4$) single crystals to a powder. The powder sample with a weight of 5–10~g/cm$^2$ according to the iron content was pressed in aluminum foil 20~mm in diameter. The processing occurred in two stages. At the first stage, possible nonequivalent positions of iron in the samples were determined by calculating the hyperfine field probability distributions. Basing on the results obtained, a preliminary model spectrum was formed. At the next stage, the model spectrum was fitted to the experimental spectrum by varying the entire set of hyperfine parameters using the least squares method in the linear approximation.

\begin{figure}[tb!]
	\centering\includegraphics[width=1\linewidth]{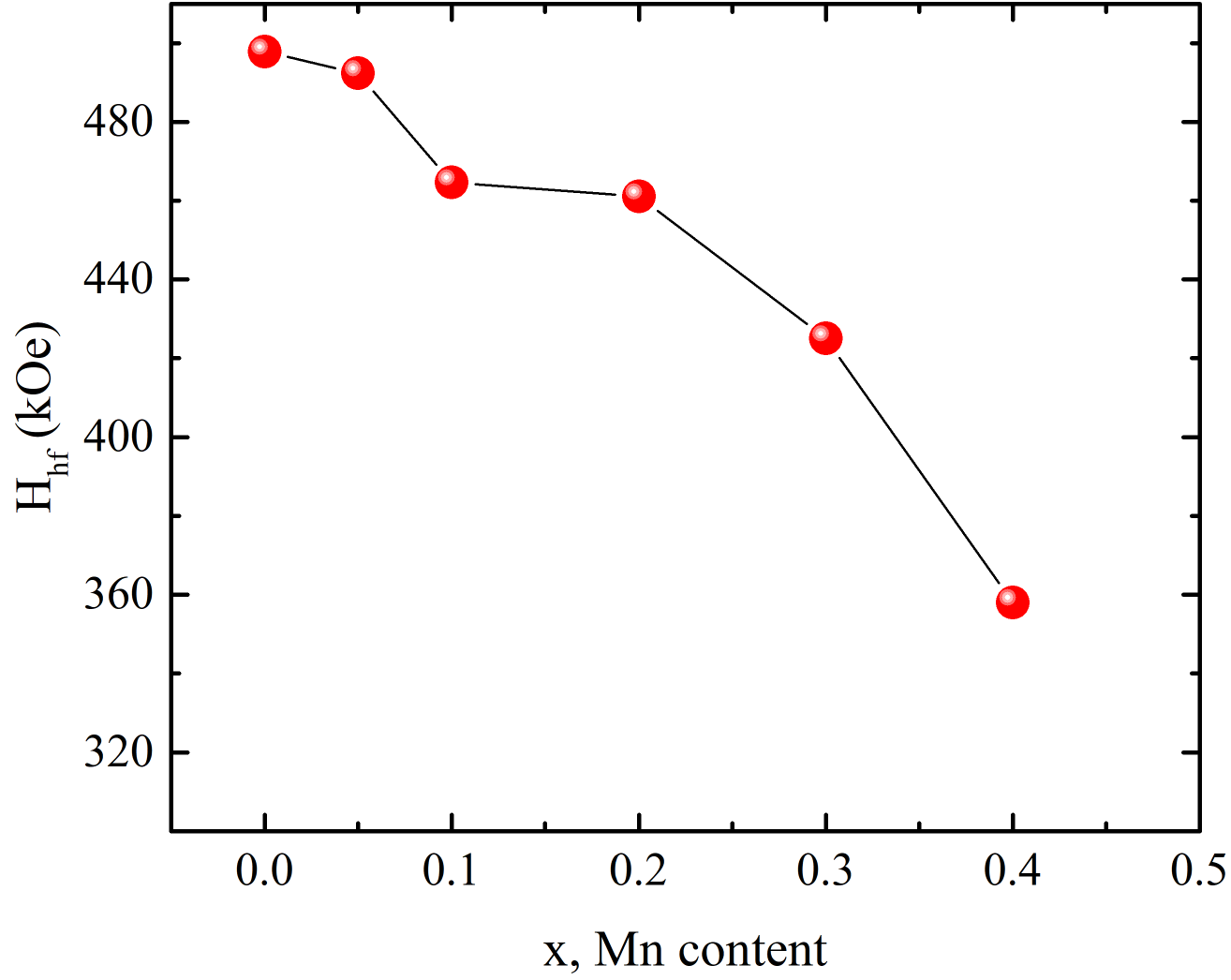}	
	\caption{~Concentration dependence of the average hyperfine field on iron nuclei in the HoFe$_{1\textendash x}$Mn$_x$O$_3$ compound.}
 \label{fig5} 
\end{figure}

\begin{figure}[tb!]
	\centering\includegraphics[width=1\linewidth]{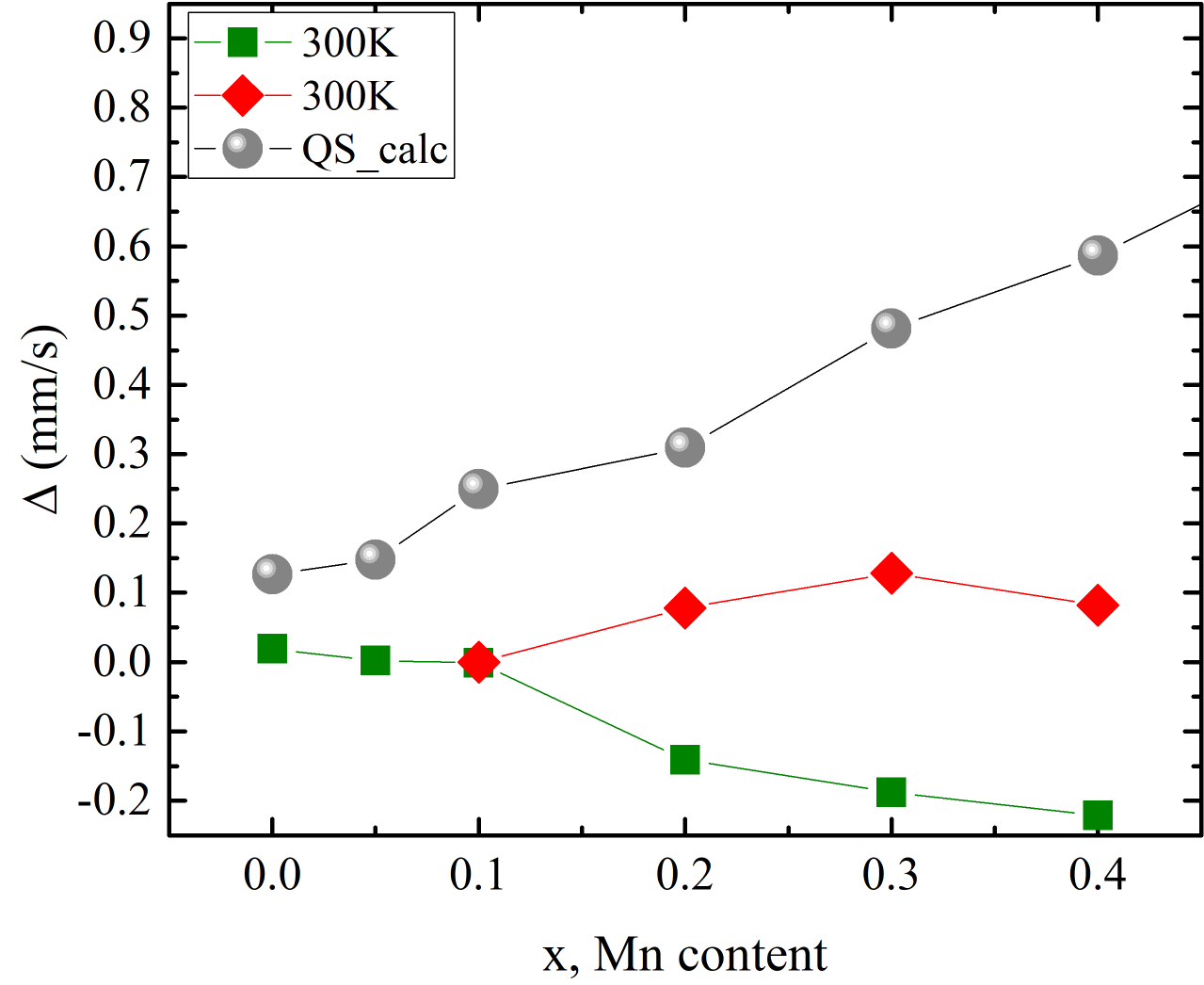}	
	\caption{~Concentration dependence of quadrupole shift $\triangle$ in the HoFe$_{1\textendash x}$Mn$_x$O$_3$ ($0<x<0.4$) samples. Spheres show the QS values calculated from the X\textendash ray diffraction data.}
 \label{fig6} 
\end{figure}

The spectra obtained are shown in Fig.~\ref{fig5}. These are fully resolved Zeeman sextets, the parameters of which are listed in Table~\ref{table3}. The chemical shift $\delta$ for all the samples relative to $\alpha$\textendash Fe indicates the $3+$ charge state of high-spin iron cations in the octahedral environment, which is consistent with the crystal structure of orthoferrites and the previous Mössbauer data~\cite{bib32,bib33}. Under the increasing substitution of Mn$^{3+}$ cations up to $x=0.20$, an additional iron site appears, which corresponds to a sextet with a strong hyperfine field. The fraction of this sextet increases stepwise and remains almost invariable in the concentration range of $x=0.20\textendash 0.40$, which may indicate that manganese cations occupy certain positions in the lattice at these concentrations. This is indirectly confirmed by two characteristic segments in the concentration dependence of the average hyperfine field on iron nuclei, which remains monotonic (Fig.~\ref{fig6}).

\begin{table}[tb!]
\caption {~Mössbauer parameters at 300~K. $\delta$ is the isomer chemical shift relative to $\alpha$\textendash Fe, H$_{hf}$ is the hyperfine field on iron nuclei, $\triangle$ is the quadrupole splitting, $W$ is the Mössbauer line full width at half maximum, $dH$ is the Mössbauer line broadening due to the inhomogeneity of the magnetic environment, and $A$ is the relative site occupancy.}
\label{table3}
\begin{ruledtabular}
\begin{tabular}{lllllll}
 & $\delta,$ & H$_{hf},$ & $\triangle,$ & $W,$ & $dH,$ & $A,$\\
 & $\pm 0.005$ & $\pm 5$ & $\pm 0.02$ & $\pm 0.03$ & $\pm 0.03$ & $\pm 0.03$\\
 & $mm/s$ & $kOe$ & $mm/s$ & $mm/s$ & $mm/s$ & $arb. u.$\\
\hline
 HoFeO$_3$\\
 $S1$ & $0.387$ & $498$ & $0.02$ & $0.26$ & $0$ & $1.0$\\
 $x=0.05$\\
 $S1$ & $0.374$ & $492$ & $0.00$ & $0.20$ & $0.20$ & $1.0$\\
 $x=0.10$\\
 $S1$ & $0.382$ & $465$ & $-0.00$ & $0.46$ & $0$ & $1.0$\\
 $x=0.20$\\
 $S1$ & $0.402$ & $452$ & $-0.14$ & $0.25$ & $0.54$ & $0.73$\\
 $S2$ & $0.388$ & $475$ & $0.08$ & $0.33$ & $0$ & $0.27$\\
 $x=0.30$\\
 $S1$ & $0.368$ & $409$ & $-0.19$ & $0.21$ & $0.73$ & $0.70$\\
 $S2$ & $0.388$ & $441$ & $0.12$ & $0.51$ & $0$ & $0.30$\\
 $x=0.40$\\
 $S1$ & $0.427$ & $335$ & $-0.20$ & $0.37$ & $0.98$ & $0.75$\\
 $S2$ & $0.361$ & $383$ & $0.08$ & $0.31$ & $0.58$ & $0.25$\\
\end{tabular}
\end{ruledtabular}
\end{table}

Let us focus on the dependence of quadrupole shift $\triangle$ of the main sextet on the manganese content in the HoFe$_{1\textendash x}$Mn$_x$O$_3$ single crystal (Fig.~\ref{fig7}). We can see a monotonic decrease in the quadrupole shift with the increasing Mn content in the samples. Simultaneously, we note the change of the $\triangle$ sign at a Mn content of $x=0.20$ for the sextet with a larger area. This can be explained by the effect of the single\textendash ion anisotropy of manganese cations, which essentially contributes to the formation of the EFG tensor. As for the behavior of the second sextet quadrupole shift, its sign remains invariable over the entire concentration range and its value changes insignificantly with an increase in the degree of substitution.

The quadrupole shift of the Mössbauer spectrum originates from (i) local distortions of the crystal lattice and (ii) the mutual direction of vectors $V_{zz}$ and $H_{hf}$. The former determine the lattice contribution to the quadrupole splitting in the paramagnetic state, which can be found, with allowance for the Sternheimer antiscreening effect~\cite{bib35,bib36} as 

\begin{eqnarray}
   QS=(1-\gamma_\infty) \frac{1}{2} e Q V_{zz} (1+\frac{\eta^2}{3})^{1/2}
    \label{Eq:eq2}
\end{eqnarray}

Here, $\gamma_\infty=‒9.44$ is the Sternheimer antiscreening factor for a spherically symmetric Fe$^{3+}$ cation~\cite{bib37} and $\eta=(V_{xx}-V_{yy})/V_{zz}$ is the a symmetry parameter, which describes the deviation from the axial symmetry.

In our case, the change in local distortions upon substitution is taken into account in the $V_{zz}$ value calculated from the X\textendash ray diffraction data. It should be noted that the approach used ignores the covalent contribution of chemical bond electrons to the quadrupole splitting. In our case, however, this contribution for the Fe$^{3+}$ cation can be neglected. The calculated data are shown by spheres in Fig.~\ref{fig7}. For the HoFeO$_3$ sample, the calculated $\triangle$ value is 0.13~mm/s, while the experimental value in the paramagnetic state is 0.30~mm/s, as reported ~\cite{bib32}. We attribute the obtained discrepancy with the experiment to the valence contribution.

\subsection{Magnetic Measurements}

\begin{figure*}[tb!]
	\centering\includegraphics[width=1\linewidth]{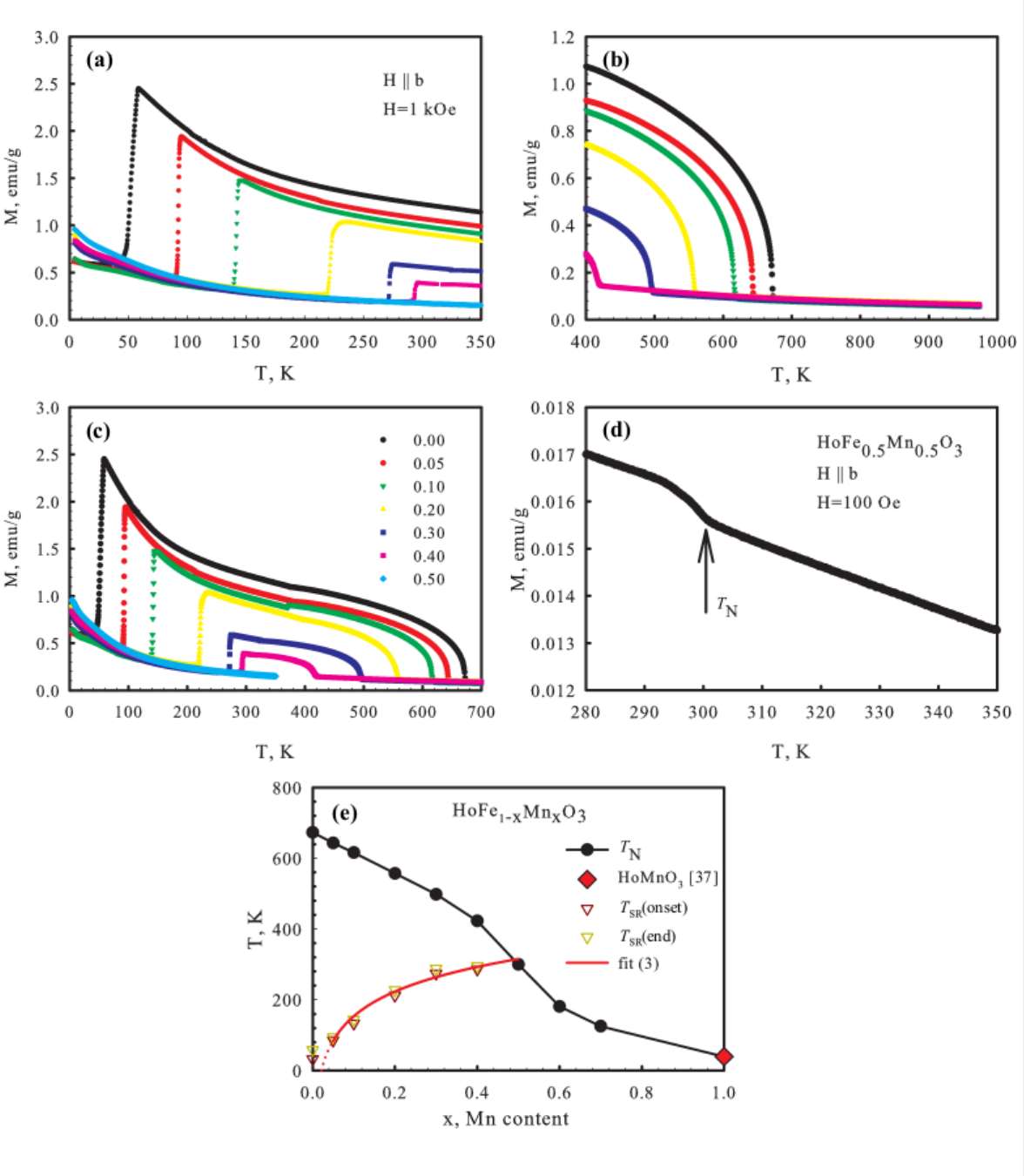}
    \caption{~Temperature dependences of magnetization $M$ of the HoFe$_{1\textendash x}$Mn$_x$O$_3$ single crystals measured in an external magnetic field of $H = 1$~kOe parallel to the b crystal axis at temperatures of (a) $4.2–350$ and (b) $400–1000$~K and (c) over the entire temperature range of $4.2–1000$~K. (d) Example of the $M(T)$ dependence for the HoFe$_{0.5}$Mn$_{0.5}$O$_3$ sample for determining the Néel temperature \TN~at high Mn contents. (e) HoFe$_{1\textendash x}$Mn$_x$O$_3$ magnetic phase diagram showing the N\'eel temperatures \TN~and the temperatures of the onset and end of the spin-orientation transition as functions of the Mn content. The red solid line is fitting of the \TSR(x) dependence by Eq.~\ref{Eq:eq3}}
    \label{fig7} 
\end{figure*}

To elucidate the effect of manganese substitution on the magnetic properties of the HoFe$_{1\textendash x}$Mn$_x$O$_3$ compounds, the temperature and field dependences of the magnetization M were measured. Figure 8 shows the $M(T)$ dependences measured in the magnetic field $H||\textbf{b}$. It can be seen that, as the manganese content increases, the temperature \TSR~shifts monotonically to the high-temperature region. For the unsubstituted HoFeO$_3$ sample, we have \TSR$\approx 58$~K, while for the sample with $x=0.4$, \TSR$=294$~K; i.e., the \TSR~values lie in the room-temperature region. The value of magnetization ($M$), which is determined by a weak ferromagnetic moment induced by canting of the iron antiferromagnetic sublattices, decreases gradually as the Mn content grows (Fig.~\ref{fig7}a).

In addition, with an increase in the manganese content, the HoFe$_{1\textendash x}$Mn$_x$O$_3$ Néel temperature decreases monotonically from \TN$=672$~K for the HoFeO$_3$ compound to \TN$=125$~K for the composition with $x=0.7$ ({Figs.~\ref{fi7} (b, c, e). An example of determining the Néel temperature of the samples with $x=0.5, 0.6,$~and~$0.7$ is presented in Fig.~\ref{fig7}d; in Fig.~\ref{fig7}d, at $x=0.5$, the Néel temperature is \TN$=300$~K. The Néel temperatures of all the investigated HoFe$_{1\textendash x}$Mn$_x$O$_3$ samples are shown in the phase diagram in Fig.~\ref{fig7}e. The \TN~value at $x=1$ (HoMnO$_3$) was determined in~\cite{bib38}. It can be seen that, at concentrations of $x\approx 0.5$, the \TN(x) dependence has an inflection, which is possibly related to a change in the dominant antiferromagnetic interaction from Fe\textendash O\textendash Fe to Mn\textendash O\textendash Mn. In addition, $x=0.4\textendash 0.5$ is the limiting concentration range, in which the spin-reorientation transition still occurs; with a further increase in the manganese content, \TN~becomes lower than \TSR, which corresponds to the paramagnet\textendash to\textendash antiferromagnet phase transition without successive orientational transitions. Figure~\ref{fig7}e shows also the dependence of the temperature \TSR~on the Mn content $x$ (red solid line) calculated using the~\cite{bib39}

\begin{eqnarray}
   T(x)=\frac{1}{k'}log(\frac{x}{x_c})
    \label{Eq:eq3}
\end{eqnarray}

where $k^{'}$ is a positive constant related to the second\textendash order anisotropy fields in the $b\textendash a$, $b\textendash c$, $c\textendash a$ planes. $x_c$ characterizes a critical doping concentration from a hypothesis that at $x = x_c$. In this study, the change in the second\textendash order anisotropy constants responsible for the temperature of the spin\textendash reorientation transition upon variation in the cobalt concentration in erbium, holmium, and dysprosium orthoferrites was investigated. In our opinion, this formalism~\cite{bib39} can be applied to describing the \TSR(x) change in the HoFe$_{1\textendash x}$Mn$_x$O$_3$ compound. Previously, formula (\ref{Eq:eq3}) was successfully used in~\cite{bib18,bib19,bib20}.

It is known well that, during the spin\textendash reorientation transition in the unsubstituted HoFeO$_3$ compound, the weak ferromagnetic moment rotates spontaneously by 90° from the \textbf{b} to \textbf{c} direction in the crystal (\textit{Pnma}) from the $A_xF_yG_z$ to $C_xG_yF_z $ phase (Fig.~\ref{fig8}); this is a second-order phase transition~\cite{bib3}. The temperature transition width for the HoFeO$_3$ composition is $\sim10$~K and no temperature hysteresis is observed. In the substituted samples, we have a completely different picture.

\begin{figure}[tb!]
	\centering\includegraphics[width=1\linewidth]{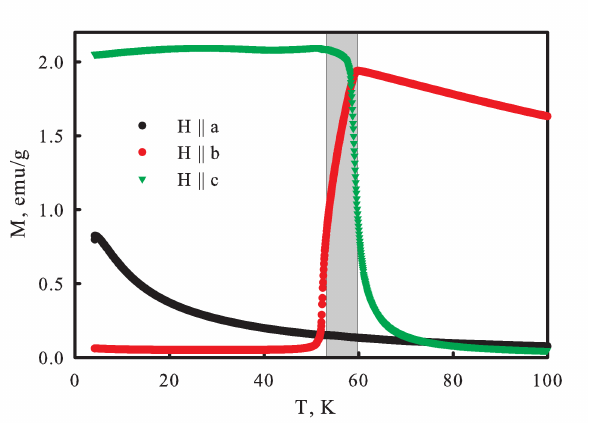}
	\caption{~Temperature dependences of magnetization M of the HoFeO$_3$ compound measured in a field of $H=100$~Oe along different crystallographic axes.  The spin reorientation area is marked in gray.}
 \label{fig8} 
\end{figure}

First, as can be seen from the $M(T)$ dependences measured in the heating mode (Figs.~\ref{fig7} (a and c)), the width of the spin-reorientation transition in the substituted samples is about $4–5$~K. Figure~\ref{fig9} shows the $M(T)$ dependences for the HoFe$_{0.7}$Mn$_{0.3}$O$_3$ sample measured in external magnetic fields of different values and configurations. According to Fig.~\ref{fig9}a, the nature of the spin\textendash reorientation transition in the substituted sample changed drastically: in the unsubstituted HoFeO$_3$ compound, the phase transition from the $A_xF_yG_z$ to $C_xG_yF_z $ phase occurs with a change in the weak ferromagnetic moment direction, while in the HoFe$_{0.7}$Mn$_{0.3}$O$_3$ compound, below the \TSR~temperature, a compensated antiferromagnetic phase of the iron and manganese sublatticeis observed. In this case, the magnetization is only determined by the paramagnetic anisotropic contribution of holmium, which is reflected in the $M(T)$ dependences (Fig.~\ref{fig9}a) measured along different crystallographic axes. In addition, it is noteworthy that, upon temperature cycling, the $M(T)$ dependences measured along the \textbf{b} direction in different magnetic fields (Fig.~\ref{fig9}b}) exhibit the hysteresis in the region of the transition (\TSR), the width of which amounts to $\sim5$~K regardless of the applied magnetic field (up to 5~kOe). In addition, it can be seen in Fig.~\ref{fig9}b that the applied magnetic field affects weakly the \TSR~temperature position; the onset of the transition shifts by 2~K toward lower temperatures in a field of $H=5$~kOe and the transition is not blurred. Exactly the same $M(T)$ behavior is exhibited by all the substituted HoFe$_{1\textendash x}$Mn$_x$O$_3$ samples, which only have different spin-reorientation transition temperature. 

\begin{figure}[tb!]
	\centering\includegraphics[width=0.975\linewidth]{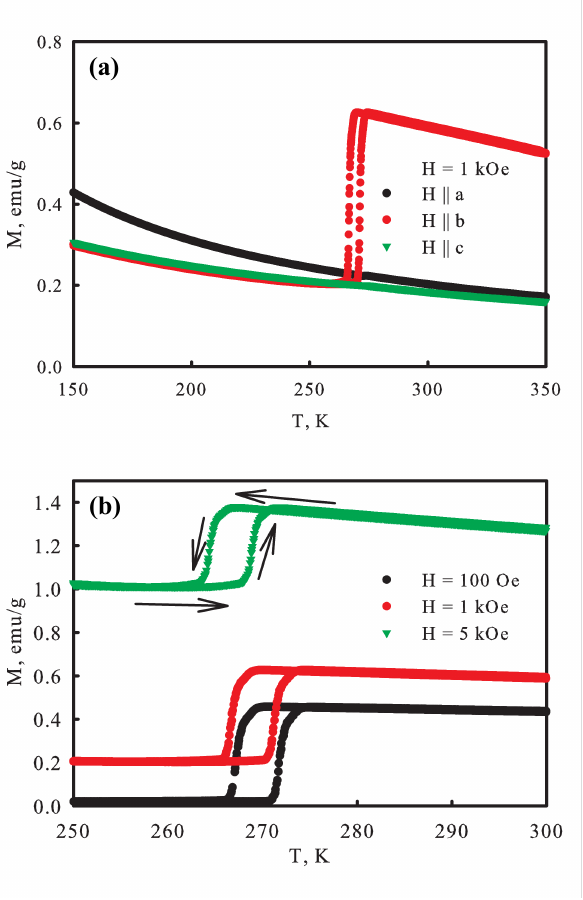}	
	\caption{~(a) Temperature dependences of magnetization M of the HoFe$_{0.7}$Mn$_{0.3}$O$_3$single crystal in the region of the spin\textendash reorientation transition. (b) Temperature dependences of magnetization $M$ measured in different applied magnetic fields $H||\textbf{b}$ in the cooling and heating modes.}
 \label{fig9} 
\end{figure}

\begin{figure}[tb!]
	\centering\includegraphics[width=1\linewidth]{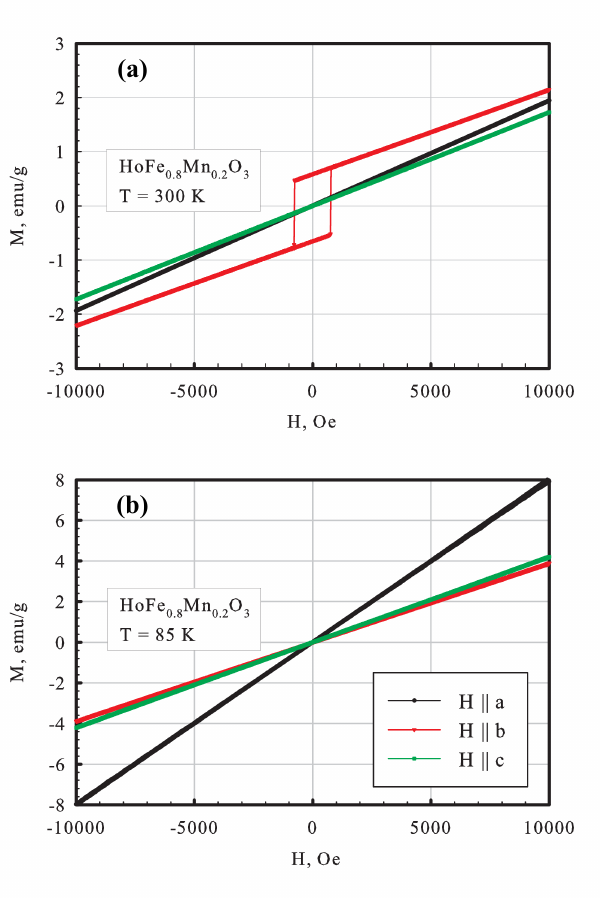}	
	\caption{~$M(H)$ dependences for the HoFe$_{0.8}$Mn$_{0.2}$O$_3$ single crystal at $T=300$ and $85$~K.}
 \label{fig10} 
\end{figure}

Figure~\ref{fig10} shows the field dependences of magnetization $M$ of the HoFe$_{0.8}$Mn$_{0.2}$O$_3$ sample at temperatures above and below the temperature \TSR~of the spin\textendash reorientation transition along three crystallographic axes. It can be seen that the $M(H)$ ferromagnetic hysteresis is only observed along the \textbf{b} direction at $T>$ \TSR, which corresponds to the $A_xF_yG_z$ magnetic phase (Fig.~\ref{fig10}a). At temperatures of $T<$ \TSR, all the $M(H)$ dependences are linear, which corresponds to the contribution of the collinear antiferromagnetic structure (the $G_xC_yA_z$ phase) without ferromagnetic component. The similar $M(H)$ behavior was observed by us for all the substituted samples.

\begin{figure*}[tb!]
	\centering\includegraphics[width=1\linewidth]{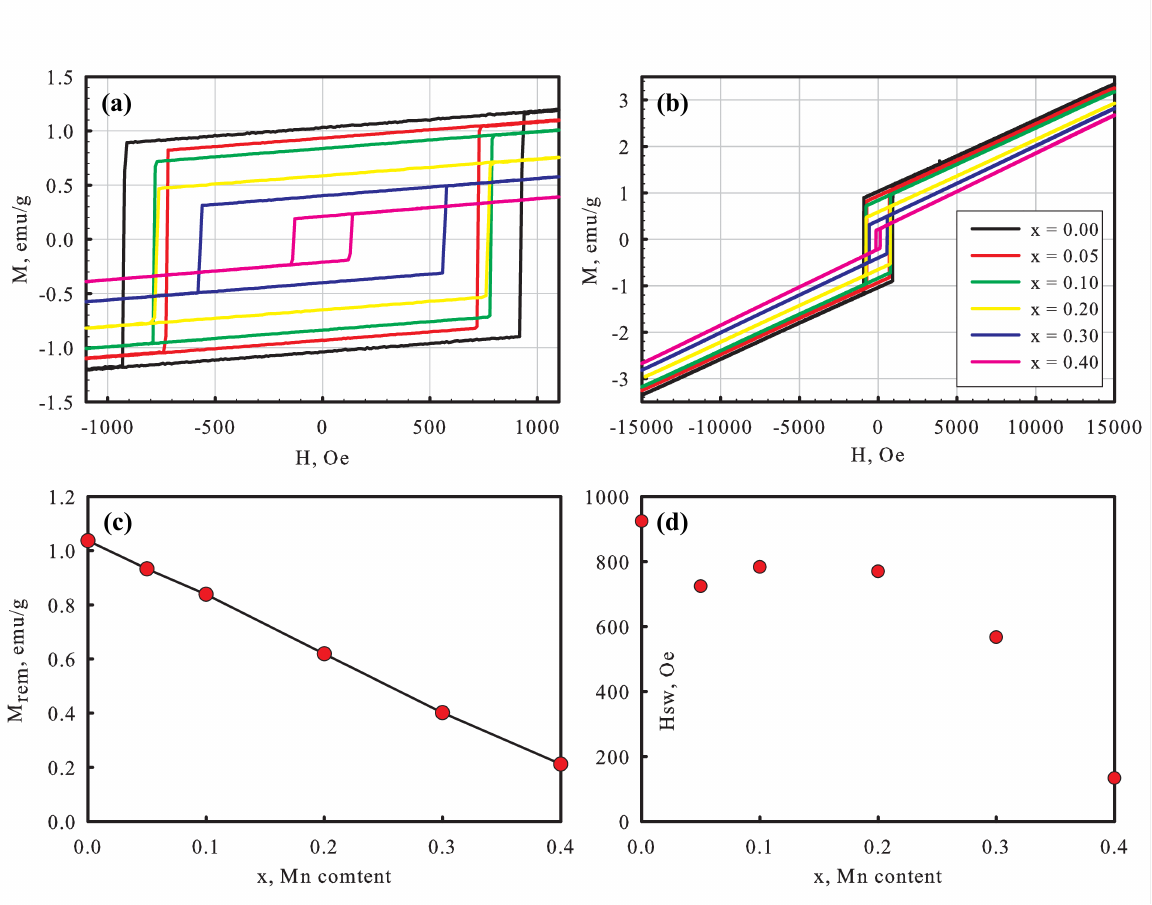}	
	\caption{~(a, b) $M(H)$ dependences at $T=300$~K for a series of the HoFe$_{1\textendash x}$Mn$_x$O$_3$ samples. (c) Remanent magnetization at $H=0$ and (d) switching field as functions of themanganese content.}
 \label{fig11} 
\end{figure*}

Figure\ref{fig11} shows the $M(H)$ dependences measured at $T=300$~K for all the HoFe$_{1\textendash x}$Mn$_x$O$_3$ samples and the remanent magnetizations at $H=0$~Oe and coercivities (switching fields) H$_{sw}$. The magnetization switching from the lower to upper branch occurs in a negligibly narrow field range, which means that the sample includes a single magnetic domain. In all the substituted samples at $T=300$~K, the ferromagnetic magnetization component only exists along the \textbf{b} direction. Therefore, by changing the Mn content in the samples, one can smoothly change both the magnetization and coercivity value, which can be promising for use of this series of single crystals at room temperatures.

\section{Discussion}
\label{discuss}

It can be concluded from the analysis of the $M(T)$ dependences that, in manganese\textendash substituted holmium orthoferrites, the spin\textendash reorientation transition is a first-order phase transition~\cite{bib3}, which determines the change in the magnetic state from the weak ferromagnetism region (the $A_xF_yG_z$ phase, the magnetic moment along the \textbf{b} crystal direction) to the fully compensated antiferromagnetic region (the $G_xC_yA_z$ phase). Such a change in the nature of the phase transition in the HoFe$_{1\textendash x}$Mn$_x$O$_3$ compound was observed in~\cite{bib28}, where this conclusion was made on the basis of the neutron powder diffraction data. The change in the magnetic transition configuration from $A_xF_yG_z\rightarrow C_xG_yF_z$ to $A_xF_yG_z\rightarrow G_xC_yA_z$ (in \textit{Pnma} notation) was observed in the TbFe$_{1\textendash x}$Mn$_x$O$_3$ compound in~\cite{bib18}, where the type of magnetic ordering was also determined by neutron diffraction. In this work, we obtained the similar results by the thorough analysis of the temperature and field dependences of the magnetization. Meanwhile, some questions about the magnetic behavior of the investigated system remain open.

Until now, no attention has been paid to the origin and mechanism of the significant growth of the temperature of the magnetic spin-reorientation transition in the HoFe$_{1\textendash x}$Mn$_x$O$_3$ system. The reasons for this phenomenon can be analyzed using the Mössbauer spectroscopy data. Since the replacement of a part of iron cations results in a change in the orbital momentum at the sites, the substitution will also change the spin‒orbit coupling value. As a result, the magnetic moment direction will deviate from the quantization axis. This will lead to a change in the mutual orientation of the EFG ($V_{zz}$) vector and the hyperfine field vector ($H_{hf}$) on iron nuclei. In the case of the axial symmetry, this change is described by the dependence of the quadrupole shift~\cite{bib31,bib34}

\begin{eqnarray}
   \triangle E_Q = \frac{e^2}{4} Q\cdot V_{zz} \frac{1}{2} (3cos^2 (\theta) - 1)
    \label{Eq:eq4}
\end{eqnarray}

Here, $Q$ is the quadrupole moment of a nucleus ($+0.21\cdot 10^{‒24}$~sm$^{‒2}$), $V_{zz}$  is the main component of the EFG tensor, $\theta$ is the angle between the EFG direction and the hyperfine field, and $e$ is the elementary charge. It is reliably demonstrated by the magnetic measurement data that the temperature of the spin-reorientation transition in the HoFe$_{1\textendash x}$Mn$_x$O$_3$ compound depends strongly on the degree of substitution. Since the crystal structure is preserved in this case, it can be assumed that the quadrupole moment of a nucleus does not change. This allows us to estimate the concentration dependence of the angle $\theta$ upon substitution using the Mössbauer spectroscopy data and Eq.~\ref{Eq:eq3}. 

\begin{figure}[tb!]
	\centering\includegraphics[width=1\linewidth]{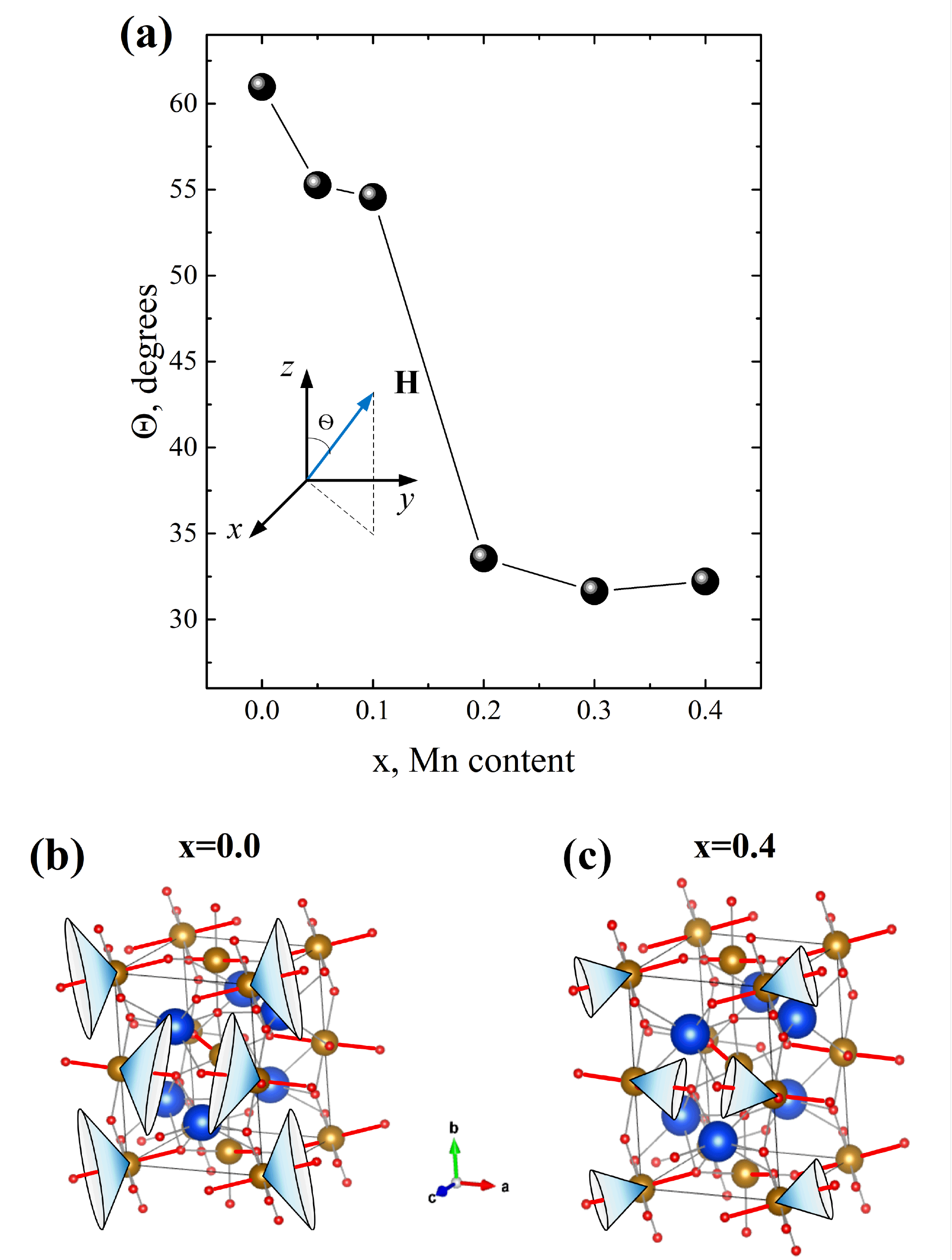}	
	\caption{~(a) Dependence of the angle of mutual arrangement of the vectors of the main EFG component and hyperfine field on iron nuclei. (b) Schematic arrangement of the hyperfine field vectors (over the cone surface) and vector $V_{zz}$ (bold line) at two extreme concentrations.}
 \label{fig12} 
\end{figure}

Thus, taking into account the $V_{zz}$ values obtained by X\textendash ray diffraction, we can estimate the change in the mutual orientation of the vectors $V_{zz}$ and $H_{hf}$ under the substitution (Fig.~\ref{fig12}). It can be clearly seen that the angle changes sharply at a Mn content of $x=0.20$. Taking into account the known direction of the vector $V_{zz}$ from Fig.~\ref{fig4}, we can demonstrate that the magnetic moment direction (opposite to $H_{hf}$) becomes closer to the \textbf{ac} crystal plane at the incorporation of manganese (Fig.~\ref{fig8}b). In~\cite{bib18}, a similar substitution in the TbFe$_{1\textendash x}$Mn$_x$O$_3$ samples was discussed. Analyzing the changes in the indirect exchange in this system, the authors noted that the Mn substitution changes the magnetic moment orbital component, which is responsible for the magnetic structure formation. According to the Mössbauer data obtained, we can conclude that the change in the mutual orientation of the vectors $V_{zz}$ and $H_{hf}$  upon substitution results from the Jahn\textendash Teller effect. This leads to the rotation of the magnetic moment both on Mn cations and on Fe cations and to a decrease in the weak ferromagnetic moment along the \textbf{b} crystal direction. This rotation of the magnetic moments in the subsystem of 3d cations in the HoFe$_{1\textendash x}$Mn$_x$O$_3$ system can explain the growth of the temperature of the spin-reorientation transition.

\

\section{Conclusion}

In this study, a series of HoFe$_{1\textendash x}$Mn$_x$O$_3$ single crystals over the entire substitute (Mn) concentration range was synthesized by the optical floating zone technique. It was found that, in the concentration range of $0.7<x<0.8$, the HoFe$_{1\textendash x}$Mn$_x$O$_3$ compound undergoes a structural transition from the rhombic ($x<0.7$) to hexagonal modification.

For the rhombic HoFe$_{1\textendash x}$Mn$_x$O$_3$ modification, the magnetic measurements were performed, which showed that the spin-reorientation transition temperature \TSR~increases significantly as the manganese content grows and, in the sample with $x=0.4$, attains room temperature, which is important for application of the substituted orthoferrites.

It was found from the analysis of the magnetic measurement data that in all the substituted samples, the spin\textendash reorientation transition is a first\textendash order phase transition from the $A_xF_yG_z\rightarrow G_xC_yA_z$ phase with decreasing temperature, whereas in the initial HoFeO$_3$ sample, this is a second-order transition from the $A_xF_yG_z\rightarrow C_xG_yF_z$ phase.

An increase in the temperature of the spin\textendash reorientation transition was attributed to a decrease in the value of the indirect Fe\textendash O\textendash Fe exchange under substitution of manganese for iron, which was established in the Mössbauer spectroscopy experiments.

\section*{Acknowledgments}

This study was supported by the Russian Science Foundation, project no. 23\textendash 22\textendash 10026, https://rscf.ru/project/23-22-10026/ and the Krasnoyarsk Territorial Foundation for Support of Scientific and R\&D Activities. The investigation of magnetic measurements and structural properties were performed using equipment from the Center for Collective Use, Krasnoyarsk Scientific Center, Siberian Branch of the Russian Academy of Sciences.

%\pagebreak
\bibliography{HoFeMnO3}

\begin{thebibliography}{39}%
\makeatletter
\providecommand \@ifxundefined [1]{%
 \@ifx{#1\undefined}
}%
\providecommand \@ifnum [1]{%
 \ifnum #1\expandafter \@firstoftwo
 \else \expandafter \@secondoftwo
 \fi
}%
\providecommand \@ifx [1]{%
 \ifx #1\expandafter \@firstoftwo
 \else \expandafter \@secondoftwo
 \fi
}%
\providecommand \natexlab [1]{#1}%
\providecommand \enquote  [1]{``#1''}%
\providecommand \bibnamefont  [1]{#1}%
\providecommand \bibfnamefont [1]{#1}%
\providecommand \citenamefont [1]{#1}%
\providecommand \href@noop [0]{\@secondoftwo}%
\providecommand \href [0]{\begingroup \@sanitize@url \@href}%
\providecommand \@href[1]{\@@startlink{#1}\@@href}%
\providecommand \@@href[1]{\endgroup#1\@@endlink}%
\providecommand \@sanitize@url [0]{\catcode `\\12\catcode `\$12\catcode
  `\&12\catcode `\#12\catcode `\^12\catcode `\_12\catcode `\%12\relax}%
\providecommand \@@startlink[1]{}%
\providecommand \@@endlink[0]{}%
\providecommand \url  [0]{\begingroup\@sanitize@url \@url }%
\providecommand \@url [1]{\endgroup\@href {#1}{\urlprefix }}%
\providecommand \urlprefix  [0]{URL }%
\providecommand \Eprint [0]{\href }%
\providecommand \doibase [0]{https://doi.org/}%
\providecommand \selectlanguage [0]{\@gobble}%
\providecommand \bibinfo  [0]{\@secondoftwo}%
\providecommand \bibfield  [0]{\@secondoftwo}%
\providecommand \translation [1]{[#1]}%
\providecommand \BibitemOpen [0]{}%
\providecommand \bibitemStop [0]{}%
\providecommand \bibitemNoStop [0]{.\EOS\space}%
\providecommand \EOS [0]{\spacefactor3000\relax}%
\providecommand \BibitemShut  [1]{\csname bibitem#1\endcsname}%
\let\auto@bib@innerbib\@empty
%</preamble>
\bibitem [{\citenamefont {White}(1969)}]{bib1}%
  \BibitemOpen
  \bibfield  {author} {\bibinfo {author} {\bibfnamefont {R.}~\bibnamefont
  {White}},\ }\bibfield  {title} {\bibinfo {title} {Review of recent work on
  the magnetic and spectroscopic properties of the rare-earth orthoferrites},\
  }\href@noop {} {\bibfield  {journal} {\bibinfo  {journal} {Journal of Applied
  Physics}\ }\textbf {\bibinfo {volume} {40}},\ \bibinfo {pages} {1061}
  (\bibinfo {year} {1969})}\BibitemShut {NoStop}%
\bibitem [{\citenamefont {Belov}\ \emph {et~al.}(1974)\citenamefont {Belov},
  \citenamefont {Zvezdin},\ and\ \citenamefont {Kadomtseva}}]{bib2}%
  \BibitemOpen
  \bibfield  {author} {\bibinfo {author} {\bibfnamefont {K.}~\bibnamefont
  {Belov}}, \bibinfo {author} {\bibfnamefont {A.}~\bibnamefont {Zvezdin}},\
  and\ \bibinfo {author} {\bibfnamefont {A.}~\bibnamefont {Kadomtseva}},\
  }\bibfield  {title} {\bibinfo {title} {New orientational transitions induced
  in orthoferrites by an external field},\ }\href@noop {} {\bibfield  {journal}
  {\bibinfo  {journal} {J. Exp. Theor. Phys}\ }\textbf {\bibinfo {volume} {67}}
  (\bibinfo {year} {1974})}\BibitemShut {NoStop}%
\bibitem [{\citenamefont {Belov}\ \emph
  {et~al.}(1979{\natexlab{a}})\citenamefont {Belov}, \citenamefont {Zvezdin},
  \citenamefont {Kadomtseva},\ and\ \citenamefont {Levitin}}]{bib3}%
  \BibitemOpen
  \bibfield  {author} {\bibinfo {author} {\bibfnamefont {K.}~\bibnamefont
  {Belov}}, \bibinfo {author} {\bibfnamefont {A.}~\bibnamefont {Zvezdin}},
  \bibinfo {author} {\bibfnamefont {A.}~\bibnamefont {Kadomtseva}},\ and\
  \bibinfo {author} {\bibfnamefont {R.}~\bibnamefont {Levitin}},\ }\href@noop
  {} {\bibinfo {title} {Orientational transitions in rare-earth magnets in
  russian}} (\bibinfo {year} {1979}{\natexlab{a}})\BibitemShut {NoStop}%
\bibitem [{\citenamefont {Belov}\ \emph
  {et~al.}(1979{\natexlab{b}})\citenamefont {Belov}, \citenamefont {Zvezdin},\
  and\ \citenamefont {Mukhin}}]{bib4}%
  \BibitemOpen
  \bibfield  {author} {\bibinfo {author} {\bibfnamefont {K.}~\bibnamefont
  {Belov}}, \bibinfo {author} {\bibfnamefont {A.}~\bibnamefont {Zvezdin}},\
  and\ \bibinfo {author} {\bibfnamefont {A.}~\bibnamefont {Mukhin}},\
  }\bibfield  {title} {\bibinfo {title} {Magnetic phase transitions in terbium
  orthoferrite},\ }\href@noop {} {\bibfield  {journal} {\bibinfo  {journal}
  {Sov. Phys. JETP}\ }\textbf {\bibinfo {volume} {49}},\ \bibinfo {pages} {557}
  (\bibinfo {year} {1979}{\natexlab{b}})}\BibitemShut {NoStop}%
\bibitem [{\citenamefont {Tokunaga}\ \emph {et~al.}(2009)\citenamefont
  {Tokunaga}, \citenamefont {Furukawa}, \citenamefont {Sakai}, \citenamefont
  {Taguchi}, \citenamefont {Arima},\ and\ \citenamefont {Tokura}}]{bib5}%
  \BibitemOpen
  \bibfield  {author} {\bibinfo {author} {\bibfnamefont {Y.}~\bibnamefont
  {Tokunaga}}, \bibinfo {author} {\bibfnamefont {N.}~\bibnamefont {Furukawa}},
  \bibinfo {author} {\bibfnamefont {H.}~\bibnamefont {Sakai}}, \bibinfo
  {author} {\bibfnamefont {Y.}~\bibnamefont {Taguchi}}, \bibinfo {author}
  {\bibfnamefont {T.-h.}\ \bibnamefont {Arima}},\ and\ \bibinfo {author}
  {\bibfnamefont {Y.}~\bibnamefont {Tokura}},\ }\bibfield  {title} {\bibinfo
  {title} {Composite domain walls in a multiferroic perovskite ferrite},\
  }\href@noop {} {\bibfield  {journal} {\bibinfo  {journal} {Nature materials}\
  }\textbf {\bibinfo {volume} {8}},\ \bibinfo {pages} {558} (\bibinfo {year}
  {2009})}\BibitemShut {NoStop}%
\bibitem [{\citenamefont {Kimel}\ \emph {et~al.}(2005)\citenamefont {Kimel},
  \citenamefont {Kirilyuk}, \citenamefont {Usachev}, \citenamefont {Pisarev},
  \citenamefont {Balbashov},\ and\ \citenamefont {Rasing}}]{bib6}%
  \BibitemOpen
  \bibfield  {author} {\bibinfo {author} {\bibfnamefont {A.}~\bibnamefont
  {Kimel}}, \bibinfo {author} {\bibfnamefont {A.}~\bibnamefont {Kirilyuk}},
  \bibinfo {author} {\bibfnamefont {P.}~\bibnamefont {Usachev}}, \bibinfo
  {author} {\bibfnamefont {R.}~\bibnamefont {Pisarev}}, \bibinfo {author}
  {\bibfnamefont {A.}~\bibnamefont {Balbashov}},\ and\ \bibinfo {author}
  {\bibfnamefont {T.}~\bibnamefont {Rasing}},\ }\bibfield  {title} {\bibinfo
  {title} {Ultrafast non-thermal control of magnetization by instantaneous
  photomagnetic pulses},\ }\href@noop {} {\bibfield  {journal} {\bibinfo
  {journal} {Nature}\ }\textbf {\bibinfo {volume} {435}},\ \bibinfo {pages}
  {655} (\bibinfo {year} {2005})}\BibitemShut {NoStop}%
\bibitem [{\citenamefont {De~Jong}\ \emph {et~al.}(2011)\citenamefont
  {De~Jong}, \citenamefont {Kimel}, \citenamefont {Pisarev}, \citenamefont
  {Kirilyuk},\ and\ \citenamefont {Rasing}}]{bib7}%
  \BibitemOpen
  \bibfield  {author} {\bibinfo {author} {\bibfnamefont {J.}~\bibnamefont
  {De~Jong}}, \bibinfo {author} {\bibfnamefont {A.}~\bibnamefont {Kimel}},
  \bibinfo {author} {\bibfnamefont {R.}~\bibnamefont {Pisarev}}, \bibinfo
  {author} {\bibfnamefont {A.}~\bibnamefont {Kirilyuk}},\ and\ \bibinfo
  {author} {\bibfnamefont {T.}~\bibnamefont {Rasing}},\ }\bibfield  {title}
  {\bibinfo {title} {Laser-induced ultrafast spin dynamics in
  $\mathrm{ErFeO}_{3}$},\ }\href@noop {} {\bibfield  {journal} {\bibinfo
  {journal} {Physical Review B}\ }\textbf {\bibinfo {volume} {84}},\ \bibinfo
  {pages} {104421} (\bibinfo {year} {2011})}\BibitemShut {NoStop}%
\bibitem [{\citenamefont {Jiang}\ \emph {et~al.}(2013)\citenamefont {Jiang},
  \citenamefont {Jin}, \citenamefont {Song}, \citenamefont {Lin}, \citenamefont
  {Ma},\ and\ \citenamefont {Cao}}]{bib8}%
  \BibitemOpen
  \bibfield  {author} {\bibinfo {author} {\bibfnamefont {J.}~\bibnamefont
  {Jiang}}, \bibinfo {author} {\bibfnamefont {Z.}~\bibnamefont {Jin}}, \bibinfo
  {author} {\bibfnamefont {G.}~\bibnamefont {Song}}, \bibinfo {author}
  {\bibfnamefont {X.}~\bibnamefont {Lin}}, \bibinfo {author} {\bibfnamefont
  {G.}~\bibnamefont {Ma}},\ and\ \bibinfo {author} {\bibfnamefont
  {S.}~\bibnamefont {Cao}},\ }\bibfield  {title} {\bibinfo {title} {Dynamical
  spin reorientation transition in $\mathrm{NdFeO}_{3}$ single crystal observed
  with polarized terahertz time domain spectroscopy},\ }\href@noop {}
  {\bibfield  {journal} {\bibinfo  {journal} {Applied Physics Letters}\
  }\textbf {\bibinfo {volume} {103}} (\bibinfo {year} {2013})}\BibitemShut
  {NoStop}%
\bibitem [{\citenamefont {Artyukhin}\ \emph {et~al.}(2012)\citenamefont
  {Artyukhin}, \citenamefont {Mostovoy}, \citenamefont {Jensen}, \citenamefont
  {Le}, \citenamefont {Prokes}, \citenamefont {De~Paula}, \citenamefont
  {Bordallo}, \citenamefont {Maljuk}, \citenamefont {Landsgesell},
  \citenamefont {Ryll} \emph {et~al.}}]{bib9}%
  \BibitemOpen
  \bibfield  {author} {\bibinfo {author} {\bibfnamefont {S.}~\bibnamefont
  {Artyukhin}}, \bibinfo {author} {\bibfnamefont {M.}~\bibnamefont {Mostovoy}},
  \bibinfo {author} {\bibfnamefont {N.~P.}\ \bibnamefont {Jensen}}, \bibinfo
  {author} {\bibfnamefont {D.}~\bibnamefont {Le}}, \bibinfo {author}
  {\bibfnamefont {K.}~\bibnamefont {Prokes}}, \bibinfo {author} {\bibfnamefont
  {V.~G.}\ \bibnamefont {De~Paula}}, \bibinfo {author} {\bibfnamefont {H.~N.}\
  \bibnamefont {Bordallo}}, \bibinfo {author} {\bibfnamefont {A.}~\bibnamefont
  {Maljuk}}, \bibinfo {author} {\bibfnamefont {S.}~\bibnamefont {Landsgesell}},
  \bibinfo {author} {\bibfnamefont {H.}~\bibnamefont {Ryll}}, \emph {et~al.},\
  }\bibfield  {title} {\bibinfo {title} {Solitonic lattice and
  $\mathrm{Y}$ukawa forces in the rare-earth orthoferrite
  $\mathrm{TbFeO}_{3}$},\ }\href@noop {} {\bibfield  {journal} {\bibinfo
  {journal} {Nature materials}\ }\textbf {\bibinfo {volume} {11}},\ \bibinfo
  {pages} {694} (\bibinfo {year} {2012})}\BibitemShut {NoStop}%
\bibitem [{\citenamefont {Nikitin}\ \emph {et~al.}(2018)\citenamefont
  {Nikitin}, \citenamefont {Wu}, \citenamefont {Sefat}, \citenamefont
  {Shaykhutdinov}, \citenamefont {Lu}, \citenamefont {Meng}, \citenamefont
  {Pomjakushina}, \citenamefont {Conder}, \citenamefont {Ehlers}, \citenamefont
  {Lumsden} \emph {et~al.}}]{bib10}%
  \BibitemOpen
  \bibfield  {author} {\bibinfo {author} {\bibfnamefont {S.~E.}\ \bibnamefont
  {Nikitin}}, \bibinfo {author} {\bibfnamefont {L.}~\bibnamefont {Wu}},
  \bibinfo {author} {\bibfnamefont {A.~S.}\ \bibnamefont {Sefat}}, \bibinfo
  {author} {\bibfnamefont {K.~A.}\ \bibnamefont {Shaykhutdinov}}, \bibinfo
  {author} {\bibfnamefont {Z.}~\bibnamefont {Lu}}, \bibinfo {author}
  {\bibfnamefont {S.}~\bibnamefont {Meng}}, \bibinfo {author} {\bibfnamefont
  {E.~V.}\ \bibnamefont {Pomjakushina}}, \bibinfo {author} {\bibfnamefont
  {K.}~\bibnamefont {Conder}}, \bibinfo {author} {\bibfnamefont
  {G.}~\bibnamefont {Ehlers}}, \bibinfo {author} {\bibfnamefont {M.~D.}\
  \bibnamefont {Lumsden}}, \emph {et~al.},\ }\bibfield  {title} {\bibinfo
  {title} {Decoupled spin dynamics in the rare-earth orthoferrite
  $\mathrm{YbFeO}_{3}$: Evolution of magnetic excitations through the
  spin-reorientation transition},\ }\href@noop {} {\bibfield  {journal}
  {\bibinfo  {journal} {Physical Review B}\ }\textbf {\bibinfo {volume} {98}},\
  \bibinfo {pages} {064424} (\bibinfo {year} {2018})}\BibitemShut {NoStop}%
\bibitem [{\citenamefont {Saito}\ \emph {et~al.}(2001)\citenamefont {Saito},
  \citenamefont {Sato}, \citenamefont {Bhattacharjee},\ and\ \citenamefont
  {Sorai}}]{bib11}%
  \BibitemOpen
  \bibfield  {author} {\bibinfo {author} {\bibfnamefont {K.}~\bibnamefont
  {Saito}}, \bibinfo {author} {\bibfnamefont {A.}~\bibnamefont {Sato}},
  \bibinfo {author} {\bibfnamefont {A.}~\bibnamefont {Bhattacharjee}},\ and\
  \bibinfo {author} {\bibfnamefont {M.}~\bibnamefont {Sorai}},\ }\bibfield
  {title} {\bibinfo {title} {High-precision detection of the heat-capacity
  anomaly due to spin reorientation in $\mathrm{TmFeO}_{3}$ and
  $\mathrm{HoFeO}_{3}$},\ }\href@noop {} {\bibfield  {journal} {\bibinfo
  {journal} {Solid state communications}\ }\textbf {\bibinfo {volume} {120}},\
  \bibinfo {pages} {129} (\bibinfo {year} {2001})}\BibitemShut {NoStop}%
\bibitem [{\citenamefont {Ovsianikov}\ \emph {et~al.}(2022)\citenamefont
  {Ovsianikov}, \citenamefont {Usmanov}, \citenamefont {Zobkalo}, \citenamefont
  {Hutanu}, \citenamefont {Barilo}, \citenamefont {Liubachko}, \citenamefont
  {Shaykhutdinov}, \citenamefont {Terentjev}, \citenamefont {Semenov},
  \citenamefont {Chatterji} \emph {et~al.}}]{bib12}%
  \BibitemOpen
  \bibfield  {author} {\bibinfo {author} {\bibfnamefont {A.}~\bibnamefont
  {Ovsianikov}}, \bibinfo {author} {\bibfnamefont {O.}~\bibnamefont {Usmanov}},
  \bibinfo {author} {\bibfnamefont {I.}~\bibnamefont {Zobkalo}}, \bibinfo
  {author} {\bibfnamefont {V.}~\bibnamefont {Hutanu}}, \bibinfo {author}
  {\bibfnamefont {S.}~\bibnamefont {Barilo}}, \bibinfo {author} {\bibfnamefont
  {N.}~\bibnamefont {Liubachko}}, \bibinfo {author} {\bibfnamefont
  {K.}~\bibnamefont {Shaykhutdinov}}, \bibinfo {author} {\bibfnamefont {K.~Y.}\
  \bibnamefont {Terentjev}}, \bibinfo {author} {\bibfnamefont {S.}~\bibnamefont
  {Semenov}}, \bibinfo {author} {\bibfnamefont {T.}~\bibnamefont {Chatterji}},
  \emph {et~al.},\ }\bibfield  {title} {\bibinfo {title} {Magnetic phase
  diagram of $\mathrm{HoFeO}_{3}$ by neutron diffraction},\ }\href@noop {}
  {\bibfield  {journal} {\bibinfo  {journal} {Journal of magnetism and magnetic
  materials}\ }\textbf {\bibinfo {volume} {557}},\ \bibinfo {pages} {169431}
  (\bibinfo {year} {2022})}\BibitemShut {NoStop}%
\bibitem [{\citenamefont {Leake}\ \emph {et~al.}(1968)\citenamefont {Leake},
  \citenamefont {Shirane},\ and\ \citenamefont {Remeika}}]{bib13}%
  \BibitemOpen
  \bibfield  {author} {\bibinfo {author} {\bibfnamefont {J.}~\bibnamefont
  {Leake}}, \bibinfo {author} {\bibfnamefont {G.}~\bibnamefont {Shirane}},\
  and\ \bibinfo {author} {\bibfnamefont {J.}~\bibnamefont {Remeika}},\
  }\bibfield  {title} {\bibinfo {title} {The magnetic structure of thulium
  orthoferrite, $\mathrm{TmFeO}_{3}$},\ }\href@noop {} {\bibfield  {journal}
  {\bibinfo  {journal} {Solid State Communications}\ }\textbf {\bibinfo
  {volume} {6}},\ \bibinfo {pages} {15} (\bibinfo {year} {1968})}\BibitemShut
  {NoStop}%
\bibitem [{\citenamefont {Skorobogatov}\ \emph {et~al.}(2022)\citenamefont
  {Skorobogatov}, \citenamefont {Shaykhutdinov}, \citenamefont {Balaev},
  \citenamefont {Pavlovskii}, \citenamefont {Krasikov},\ and\ \citenamefont
  {Terentjev}}]{bib14}%
  \BibitemOpen
  \bibfield  {author} {\bibinfo {author} {\bibfnamefont {S.}~\bibnamefont
  {Skorobogatov}}, \bibinfo {author} {\bibfnamefont {K.}~\bibnamefont
  {Shaykhutdinov}}, \bibinfo {author} {\bibfnamefont {D.}~\bibnamefont
  {Balaev}}, \bibinfo {author} {\bibfnamefont {M.}~\bibnamefont {Pavlovskii}},
  \bibinfo {author} {\bibfnamefont {A.}~\bibnamefont {Krasikov}},\ and\
  \bibinfo {author} {\bibfnamefont {K.~Y.}\ \bibnamefont {Terentjev}},\
  }\bibfield  {title} {\bibinfo {title} {Spin dynamics and exchange interaction
  in orthoferrite $\mathrm{TbFeO}_{3}$ with non-$\mathrm{K}$ramers rare-earth
  ion},\ }\href@noop {} {\bibfield  {journal} {\bibinfo  {journal} {Physical
  Review B}\ }\textbf {\bibinfo {volume} {106}},\ \bibinfo {pages} {184404}
  (\bibinfo {year} {2022})}\BibitemShut {NoStop}%
\bibitem [{\citenamefont {Cao}\ \emph {et~al.}(2014)\citenamefont {Cao},
  \citenamefont {Zhao}, \citenamefont {Kang}, \citenamefont {Zhang},\ and\
  \citenamefont {Ren}}]{bib15}%
  \BibitemOpen
  \bibfield  {author} {\bibinfo {author} {\bibfnamefont {S.}~\bibnamefont
  {Cao}}, \bibinfo {author} {\bibfnamefont {H.}~\bibnamefont {Zhao}}, \bibinfo
  {author} {\bibfnamefont {B.}~\bibnamefont {Kang}}, \bibinfo {author}
  {\bibfnamefont {J.}~\bibnamefont {Zhang}},\ and\ \bibinfo {author}
  {\bibfnamefont {W.}~\bibnamefont {Ren}},\ }\bibfield  {title} {\bibinfo
  {title} {Temperature induced spin switching in $\mathrm{SmFeO}_{3}$ single
  crystal},\ }\href@noop {} {\bibfield  {journal} {\bibinfo  {journal}
  {Scientific reports}\ }\textbf {\bibinfo {volume} {4}},\ \bibinfo {pages}
  {5960} (\bibinfo {year} {2014})}\BibitemShut {NoStop}%
\bibitem [{\citenamefont {Dzyaloshinsky}(1958)}]{bib16}%
  \BibitemOpen
  \bibfield  {author} {\bibinfo {author} {\bibfnamefont {I.}~\bibnamefont
  {Dzyaloshinsky}},\ }\bibfield  {title} {\bibinfo {title} {A thermodynamic
  theory of “weak” ferromagnetism of antiferromagnetics},\ }\href@noop {}
  {\bibfield  {journal} {\bibinfo  {journal} {Journal of physics and chemistry
  of solids}\ }\textbf {\bibinfo {volume} {4}},\ \bibinfo {pages} {241}
  (\bibinfo {year} {1958})}\BibitemShut {NoStop}%
\bibitem [{\citenamefont {Moriya}(1960)}]{bib17}%
  \BibitemOpen
  \bibfield  {author} {\bibinfo {author} {\bibfnamefont {T.}~\bibnamefont
  {Moriya}},\ }\bibfield  {title} {\bibinfo {title} {Anisotropic superexchange
  interaction and weak ferromagnetism},\ }\href@noop {} {\bibfield  {journal}
  {\bibinfo  {journal} {Physical review}\ }\textbf {\bibinfo {volume} {120}},\
  \bibinfo {pages} {91} (\bibinfo {year} {1960})}\BibitemShut {NoStop}%
\bibitem [{\citenamefont {Fang}\ \emph {et~al.}(2016)\citenamefont {Fang},
  \citenamefont {Yang}, \citenamefont {Liu}, \citenamefont {Kang},
  \citenamefont {Hao}, \citenamefont {Chen}, \citenamefont {Xie}, \citenamefont
  {Sun}, \citenamefont {Chandragiri}, \citenamefont {Wang} \emph
  {et~al.}}]{bib18}%
  \BibitemOpen
  \bibfield  {author} {\bibinfo {author} {\bibfnamefont {Y.}~\bibnamefont
  {Fang}}, \bibinfo {author} {\bibfnamefont {Y.}~\bibnamefont {Yang}}, \bibinfo
  {author} {\bibfnamefont {X.}~\bibnamefont {Liu}}, \bibinfo {author}
  {\bibfnamefont {J.}~\bibnamefont {Kang}}, \bibinfo {author} {\bibfnamefont
  {L.}~\bibnamefont {Hao}}, \bibinfo {author} {\bibfnamefont {X.}~\bibnamefont
  {Chen}}, \bibinfo {author} {\bibfnamefont {L.}~\bibnamefont {Xie}}, \bibinfo
  {author} {\bibfnamefont {G.}~\bibnamefont {Sun}}, \bibinfo {author}
  {\bibfnamefont {V.}~\bibnamefont {Chandragiri}}, \bibinfo {author}
  {\bibfnamefont {C.-W.}\ \bibnamefont {Wang}}, \emph {et~al.},\ }\bibfield
  {title} {\bibinfo {title} {Observation of re-entrant spin reorientation in
  $\mathrm{TbFe}_{1-x}{Mn}_{x}{O}_{3}$},\ }\href@noop {} {\bibfield  {journal}
  {\bibinfo  {journal} {Scientific reports}\ }\textbf {\bibinfo {volume} {6}},\
  \bibinfo {pages} {33448} (\bibinfo {year} {2016})}\BibitemShut {NoStop}%
\bibitem [{\citenamefont {Kang}\ \emph {et~al.}(2017)\citenamefont {Kang},
  \citenamefont {Yang}, \citenamefont {Qian}, \citenamefont {Xu}, \citenamefont
  {Cui}, \citenamefont {Fang}, \citenamefont {Chandragiri}, \citenamefont
  {Kang}, \citenamefont {Chen}, \citenamefont {Stroppa} \emph
  {et~al.}}]{bib19}%
  \BibitemOpen
  \bibfield  {author} {\bibinfo {author} {\bibfnamefont {J.}~\bibnamefont
  {Kang}}, \bibinfo {author} {\bibfnamefont {Y.}~\bibnamefont {Yang}}, \bibinfo
  {author} {\bibfnamefont {X.}~\bibnamefont {Qian}}, \bibinfo {author}
  {\bibfnamefont {K.}~\bibnamefont {Xu}}, \bibinfo {author} {\bibfnamefont
  {X.}~\bibnamefont {Cui}}, \bibinfo {author} {\bibfnamefont {Y.}~\bibnamefont
  {Fang}}, \bibinfo {author} {\bibfnamefont {V.}~\bibnamefont {Chandragiri}},
  \bibinfo {author} {\bibfnamefont {B.}~\bibnamefont {Kang}}, \bibinfo {author}
  {\bibfnamefont {B.}~\bibnamefont {Chen}}, \bibinfo {author} {\bibfnamefont
  {A.}~\bibnamefont {Stroppa}}, \emph {et~al.},\ }\bibfield  {title} {\bibinfo
  {title} {Spin-reorientation magnetic transitions in $\mathrm{M}$n-doped
  $\mathrm{SmFeO}_{3}$},\ }\href@noop {} {\bibfield  {journal} {\bibinfo
  {journal} {IUCrJ}\ }\textbf {\bibinfo {volume} {4}},\ \bibinfo {pages} {598}
  (\bibinfo {year} {2017})}\BibitemShut {NoStop}%
\bibitem [{\citenamefont {Fan}\ \emph {et~al.}(2022)\citenamefont {Fan},
  \citenamefont {Chen}, \citenamefont {Zhao}, \citenamefont {Ma}, \citenamefont
  {Chakaravarthy}, \citenamefont {Kang}, \citenamefont {Lu}, \citenamefont
  {Ren}, \citenamefont {Zhang},\ and\ \citenamefont {Cao}}]{bib20}%
  \BibitemOpen
  \bibfield  {author} {\bibinfo {author} {\bibfnamefont {W.}~\bibnamefont
  {Fan}}, \bibinfo {author} {\bibfnamefont {H.}~\bibnamefont {Chen}}, \bibinfo
  {author} {\bibfnamefont {G.}~\bibnamefont {Zhao}}, \bibinfo {author}
  {\bibfnamefont {X.}~\bibnamefont {Ma}}, \bibinfo {author} {\bibfnamefont
  {R.}~\bibnamefont {Chakaravarthy}}, \bibinfo {author} {\bibfnamefont
  {B.}~\bibnamefont {Kang}}, \bibinfo {author} {\bibfnamefont {W.}~\bibnamefont
  {Lu}}, \bibinfo {author} {\bibfnamefont {W.}~\bibnamefont {Ren}}, \bibinfo
  {author} {\bibfnamefont {J.}~\bibnamefont {Zhang}},\ and\ \bibinfo {author}
  {\bibfnamefont {S.}~\bibnamefont {Cao}},\ }\bibfield  {title} {\bibinfo
  {title} {Thermal control magnetic switching dominated by spin reorientation
  transition in $\mathrm{Mn}$-doped $\mathrm{PrFeO}_{3}$ single crystals},\
  }\href@noop {} {\bibfield  {journal} {\bibinfo  {journal} {Frontiers of
  Physics}\ }\textbf {\bibinfo {volume} {17}},\ \bibinfo {pages} {1} (\bibinfo
  {year} {2022})}\BibitemShut {NoStop}%
\bibitem [{\citenamefont {Su}\ \emph {et~al.}(2019)\citenamefont {Su},
  \citenamefont {Zhang}, \citenamefont {Dong}, \citenamefont {Ke},
  \citenamefont {Hou}, \citenamefont {Yang},\ and\ \citenamefont
  {Cheng}}]{bib21}%
  \BibitemOpen
  \bibfield  {author} {\bibinfo {author} {\bibfnamefont {L.}~\bibnamefont
  {Su}}, \bibinfo {author} {\bibfnamefont {X.-Q.}\ \bibnamefont {Zhang}},
  \bibinfo {author} {\bibfnamefont {Q.-Y.}\ \bibnamefont {Dong}}, \bibinfo
  {author} {\bibfnamefont {Y.-J.}\ \bibnamefont {Ke}}, \bibinfo {author}
  {\bibfnamefont {K.-Y.}\ \bibnamefont {Hou}}, \bibinfo {author} {\bibfnamefont
  {H.-t.}\ \bibnamefont {Yang}},\ and\ \bibinfo {author} {\bibfnamefont
  {Z.-H.}\ \bibnamefont {Cheng}},\ }\bibfield  {title} {\bibinfo {title} {Spin
  reorientation and magnetocaloric effect of
  $\mathrm{GdFe}_{1-x}\mathrm{Mn}_{x}\mathrm{O}_{3}$ ($0\le x\le 0.3$) single
  crystals},\ }\href@noop {} {\bibfield  {journal} {\bibinfo  {journal}
  {Physica B: Condensed Matter}\ }\textbf {\bibinfo {volume} {575}},\ \bibinfo
  {pages} {411687} (\bibinfo {year} {2019})}\BibitemShut {NoStop}%
\bibitem [{\citenamefont {Chiang}\ \emph {et~al.}(2011)\citenamefont {Chiang},
  \citenamefont {Chu}, \citenamefont {Chou}, \citenamefont {Jeng},
  \citenamefont {Sheu}, \citenamefont {Chen},\ and\ \citenamefont
  {Chen}}]{bib22}%
  \BibitemOpen
  \bibfield  {author} {\bibinfo {author} {\bibfnamefont {F.-K.}\ \bibnamefont
  {Chiang}}, \bibinfo {author} {\bibfnamefont {M.-W.}\ \bibnamefont {Chu}},
  \bibinfo {author} {\bibfnamefont {F.}~\bibnamefont {Chou}}, \bibinfo {author}
  {\bibfnamefont {H.}~\bibnamefont {Jeng}}, \bibinfo {author} {\bibfnamefont
  {H.}~\bibnamefont {Sheu}}, \bibinfo {author} {\bibfnamefont {F.}~\bibnamefont
  {Chen}},\ and\ \bibinfo {author} {\bibfnamefont {C.}~\bibnamefont {Chen}},\
  }\bibfield  {title} {\bibinfo {title} {Effect of
  $\mathrm{J}$ahn-$\mathrm{T}$eller distortion on magnetic ordering in
  $\mathrm{Dy(FeMn)O}_{3}$ perovskites},\ }\href@noop {} {\bibfield  {journal}
  {\bibinfo  {journal} {Physical Review B}\ }\textbf {\bibinfo {volume} {83}},\
  \bibinfo {pages} {245105} (\bibinfo {year} {2011})}\BibitemShut {NoStop}%
\bibitem [{\citenamefont {Vilarinho}\ \emph {et~al.}(2018)\citenamefont
  {Vilarinho}, \citenamefont {Passos}, \citenamefont {Queir{\'o}s},
  \citenamefont {Tavares}, \citenamefont {Almeida}, \citenamefont {Weber},
  \citenamefont {Guennou}, \citenamefont {Kreisel},\ and\ \citenamefont
  {Moreira}}]{bib23}%
  \BibitemOpen
  \bibfield  {author} {\bibinfo {author} {\bibfnamefont {R.}~\bibnamefont
  {Vilarinho}}, \bibinfo {author} {\bibfnamefont {D.}~\bibnamefont {Passos}},
  \bibinfo {author} {\bibfnamefont {E.}~\bibnamefont {Queir{\'o}s}}, \bibinfo
  {author} {\bibfnamefont {P.}~\bibnamefont {Tavares}}, \bibinfo {author}
  {\bibfnamefont {A.}~\bibnamefont {Almeida}}, \bibinfo {author} {\bibfnamefont
  {M.}~\bibnamefont {Weber}}, \bibinfo {author} {\bibfnamefont
  {M.}~\bibnamefont {Guennou}}, \bibinfo {author} {\bibfnamefont
  {J.}~\bibnamefont {Kreisel}},\ and\ \bibinfo {author} {\bibfnamefont {J.~A.}\
  \bibnamefont {Moreira}},\ }\bibfield  {title} {\bibinfo {title} {Suppression
  of the cooperative $\mathrm{J}$ahn-$\mathrm{T}$eller distortion and its
  effect on the $\mathrm{R}$aman octahedra-rotation modes of
  $\mathrm{TbMn}_{1-x}\mathrm{Fe}_{x}\mathrm{O}_{3}$},\ }\href@noop {}
  {\bibfield  {journal} {\bibinfo  {journal} {Physical Review B}\ }\textbf
  {\bibinfo {volume} {97}},\ \bibinfo {pages} {144110} (\bibinfo {year}
  {2018})}\BibitemShut {NoStop}%
\bibitem [{\citenamefont {Mandal}\ \emph {et~al.}(2011)\citenamefont {Mandal},
  \citenamefont {Bhadram}, \citenamefont {Sundarayya}, \citenamefont
  {Narayana}, \citenamefont {Sundaresan},\ and\ \citenamefont {Rao}}]{bib24}%
  \BibitemOpen
  \bibfield  {author} {\bibinfo {author} {\bibfnamefont {P.}~\bibnamefont
  {Mandal}}, \bibinfo {author} {\bibfnamefont {V.~S.}\ \bibnamefont {Bhadram}},
  \bibinfo {author} {\bibfnamefont {Y.}~\bibnamefont {Sundarayya}}, \bibinfo
  {author} {\bibfnamefont {C.}~\bibnamefont {Narayana}}, \bibinfo {author}
  {\bibfnamefont {A.}~\bibnamefont {Sundaresan}},\ and\ \bibinfo {author}
  {\bibfnamefont {C.}~\bibnamefont {Rao}},\ }\bibfield  {title} {\bibinfo
  {title} {Spin-reorientation, ferroelectricity, and magnetodielectric effect
  in $\mathrm{YFe}_{1-x}\mathrm{Mn}_{x}\mathrm{O}_{3}$ ($0.1\le x\le 0.40$)},\
  }\href@noop {} {\bibfield  {journal} {\bibinfo  {journal} {Physical review
  letters}\ }\textbf {\bibinfo {volume} {107}},\ \bibinfo {pages} {137202}
  (\bibinfo {year} {2011})}\BibitemShut {NoStop}%
\bibitem [{\citenamefont {Song}\ \emph {et~al.}(2020)\citenamefont {Song},
  \citenamefont {Su}, \citenamefont {Fang}, \citenamefont {Tong}, \citenamefont
  {Xu}, \citenamefont {Yang},\ and\ \citenamefont {Zhang}}]{bib25}%
  \BibitemOpen
  \bibfield  {author} {\bibinfo {author} {\bibfnamefont {G.}~\bibnamefont
  {Song}}, \bibinfo {author} {\bibfnamefont {J.}~\bibnamefont {Su}}, \bibinfo
  {author} {\bibfnamefont {S.}~\bibnamefont {Fang}}, \bibinfo {author}
  {\bibfnamefont {J.}~\bibnamefont {Tong}}, \bibinfo {author} {\bibfnamefont
  {X.}~\bibnamefont {Xu}}, \bibinfo {author} {\bibfnamefont {H.}~\bibnamefont
  {Yang}},\ and\ \bibinfo {author} {\bibfnamefont {N.}~\bibnamefont {Zhang}},\
  }\bibfield  {title} {\bibinfo {title} {Modified crystal structure, dielectric
  and magnetic properties of cr doped $\mathrm{SmFeO}_{3}$ ceramic},\
  }\href@noop {} {\bibfield  {journal} {\bibinfo  {journal} {Physica B:
  Condensed Matter}\ }\textbf {\bibinfo {volume} {589}},\ \bibinfo {pages}
  {412185} (\bibinfo {year} {2020})}\BibitemShut {NoStop}%
\bibitem [{\citenamefont {Habib}\ \emph {et~al.}(2017)\citenamefont {Habib},
  \citenamefont {Ikram}, \citenamefont {Sultan}, \citenamefont {Abida},
  \citenamefont {Mir}, \citenamefont {Majid},\ and\ \citenamefont
  {Asokan}}]{bib26}%
  \BibitemOpen
  \bibfield  {author} {\bibinfo {author} {\bibfnamefont {Z.}~\bibnamefont
  {Habib}}, \bibinfo {author} {\bibfnamefont {M.}~\bibnamefont {Ikram}},
  \bibinfo {author} {\bibfnamefont {K.}~\bibnamefont {Sultan}}, \bibinfo
  {author} {\bibnamefont {Abida}}, \bibinfo {author} {\bibfnamefont {S.~A.}\
  \bibnamefont {Mir}}, \bibinfo {author} {\bibfnamefont {K.}~\bibnamefont
  {Majid}},\ and\ \bibinfo {author} {\bibfnamefont {K.}~\bibnamefont
  {Asokan}},\ }\bibfield  {title} {\bibinfo {title} {Electronic
  excitation-induced structural, optical, and magnetic properties of
  $\mathrm{Ni}$-doped $\mathrm{HoFeO}_{3}$ thin films},\ }\href@noop {}
  {\bibfield  {journal} {\bibinfo  {journal} {Applied Physics A}\ }\textbf
  {\bibinfo {volume} {123}},\ \bibinfo {pages} {1} (\bibinfo {year}
  {2017})}\BibitemShut {NoStop}%
\bibitem [{\citenamefont {Somvanshi}\ \emph {et~al.}(2021)\citenamefont
  {Somvanshi}, \citenamefont {Husain}, \citenamefont {Manzoor}, \citenamefont
  {Zarrin}, \citenamefont {Ahmad}, \citenamefont {Want},\ and\ \citenamefont
  {Khan}}]{bib27}%
  \BibitemOpen
  \bibfield  {author} {\bibinfo {author} {\bibfnamefont {A.}~\bibnamefont
  {Somvanshi}}, \bibinfo {author} {\bibfnamefont {S.}~\bibnamefont {Husain}},
  \bibinfo {author} {\bibfnamefont {S.}~\bibnamefont {Manzoor}}, \bibinfo
  {author} {\bibfnamefont {N.}~\bibnamefont {Zarrin}}, \bibinfo {author}
  {\bibfnamefont {N.}~\bibnamefont {Ahmad}}, \bibinfo {author} {\bibfnamefont
  {B.}~\bibnamefont {Want}},\ and\ \bibinfo {author} {\bibfnamefont
  {W.}~\bibnamefont {Khan}},\ }\bibfield  {title} {\bibinfo {title} {Tuning of
  magnetic properties and multiferroic nature: case study of cobalt-doped
  $\mathrm{NdFeO}_{3}$},\ }\href@noop {} {\bibfield  {journal} {\bibinfo
  {journal} {Applied Physics A}\ }\textbf {\bibinfo {volume} {127}},\ \bibinfo
  {pages} {1} (\bibinfo {year} {2021})}\BibitemShut {NoStop}%
\bibitem [{\citenamefont {Prakash}\ \emph {et~al.}(2019)\citenamefont
  {Prakash}, \citenamefont {Sathe}, \citenamefont {Prajapat}, \citenamefont
  {Nigam}, \citenamefont {Krishna},\ and\ \citenamefont {Das}}]{bib28}%
  \BibitemOpen
  \bibfield  {author} {\bibinfo {author} {\bibfnamefont {P.}~\bibnamefont
  {Prakash}}, \bibinfo {author} {\bibfnamefont {V.}~\bibnamefont {Sathe}},
  \bibinfo {author} {\bibfnamefont {C.}~\bibnamefont {Prajapat}}, \bibinfo
  {author} {\bibfnamefont {A.}~\bibnamefont {Nigam}}, \bibinfo {author}
  {\bibfnamefont {P.}~\bibnamefont {Krishna}},\ and\ \bibinfo {author}
  {\bibfnamefont {A.}~\bibnamefont {Das}},\ }\bibfield  {title} {\bibinfo
  {title} {Spin phonon coupling in $\mathrm{Mn}$ doped $\mathrm{HoFeO}_{3}$
  compounds exhibiting spin reorientation behaviour},\ }\href@noop {}
  {\bibfield  {journal} {\bibinfo  {journal} {Journal of Physics: Condensed
  Matter}\ }\textbf {\bibinfo {volume} {32}},\ \bibinfo {pages} {095801}
  (\bibinfo {year} {2019})}\BibitemShut {NoStop}%
\bibitem [{\citenamefont {Lee}\ \emph {et~al.}(2011)\citenamefont {Lee},
  \citenamefont {Choi}, \citenamefont {Ramazanoglu}, \citenamefont
  {W~Ratcliff}, \citenamefont {Kiryukhin},\ and\ \citenamefont
  {Cheong}}]{bib29}%
  \BibitemOpen
  \bibfield  {author} {\bibinfo {author} {\bibfnamefont {N.}~\bibnamefont
  {Lee}}, \bibinfo {author} {\bibfnamefont {Y.}~\bibnamefont {Choi}}, \bibinfo
  {author} {\bibfnamefont {M.}~\bibnamefont {Ramazanoglu}}, \bibinfo {author}
  {\bibfnamefont {I.}~\bibnamefont {W~Ratcliff}}, \bibinfo {author}
  {\bibfnamefont {V.}~\bibnamefont {Kiryukhin}},\ and\ \bibinfo {author}
  {\bibfnamefont {S.-W.}\ \bibnamefont {Cheong}},\ }\bibfield  {title}
  {\bibinfo {title} {Mechanism of exchange striction of ferroelectricity in
  multiferroic orthorhombic $\mathrm{HoMnO}_{3}$ single crystals},\ }\href@noop
  {} {\bibfield  {journal} {\bibinfo  {journal} {Physical Review B}\ }\textbf
  {\bibinfo {volume} {84}},\ \bibinfo {pages} {020101} (\bibinfo {year}
  {2011})}\BibitemShut {NoStop}%
\bibitem [{\citenamefont {Dubrovskiy}\ \emph {et~al.}(2017)\citenamefont
  {Dubrovskiy}, \citenamefont {Pavlovskiy}, \citenamefont {Semenov},
  \citenamefont {Terentjev},\ and\ \citenamefont {Shaykhutdinov}}]{bib30}%
  \BibitemOpen
  \bibfield  {author} {\bibinfo {author} {\bibfnamefont {A.}~\bibnamefont
  {Dubrovskiy}}, \bibinfo {author} {\bibfnamefont {N.}~\bibnamefont
  {Pavlovskiy}}, \bibinfo {author} {\bibfnamefont {S.}~\bibnamefont {Semenov}},
  \bibinfo {author} {\bibfnamefont {K.~Y.}\ \bibnamefont {Terentjev}},\ and\
  \bibinfo {author} {\bibfnamefont {K.}~\bibnamefont {Shaykhutdinov}},\
  }\bibfield  {title} {\bibinfo {title} {The magnetostriction of the
  $\mathrm{HoMnO}_{3}$ hexagonal single crystals},\ }\href@noop {} {\bibfield
  {journal} {\bibinfo  {journal} {Journal of Magnetism and Magnetic Materials}\
  }\textbf {\bibinfo {volume} {440}},\ \bibinfo {pages} {44} (\bibinfo {year}
  {2017})}\BibitemShut {NoStop}%
\bibitem [{\citenamefont {G{\"u}tlich}\ \emph {et~al.}(2010)\citenamefont
  {G{\"u}tlich}, \citenamefont {Bill},\ and\ \citenamefont
  {Trautwein}}]{bib31}%
  \BibitemOpen
  \bibfield  {author} {\bibinfo {author} {\bibfnamefont {P.}~\bibnamefont
  {G{\"u}tlich}}, \bibinfo {author} {\bibfnamefont {E.}~\bibnamefont {Bill}},\
  and\ \bibinfo {author} {\bibfnamefont {A.~X.}\ \bibnamefont {Trautwein}},\
  }\href@noop {} {\emph {\bibinfo {title} {M{\"o}ssbauer spectroscopy and
  transition metal chemistry: fundamentals and applications}}}\ (\bibinfo
  {publisher} {Springer Science \& Business Media},\ \bibinfo {year}
  {2010})\BibitemShut {NoStop}%
\bibitem [{\citenamefont {Angadi}\ \emph {et~al.}(2020)\citenamefont {Angadi},
  \citenamefont {Manjunatha}, \citenamefont {Kubrin}, \citenamefont {Kozakov},
  \citenamefont {Kochur}, \citenamefont {Nikolskii}, \citenamefont {Petrov},
  \citenamefont {Shevtsova},\ and\ \citenamefont {Ayachit}}]{bib32}%
  \BibitemOpen
  \bibfield  {author} {\bibinfo {author} {\bibfnamefont {V.~J.}\ \bibnamefont
  {Angadi}}, \bibinfo {author} {\bibfnamefont {K.}~\bibnamefont {Manjunatha}},
  \bibinfo {author} {\bibfnamefont {S.}~\bibnamefont {Kubrin}}, \bibinfo
  {author} {\bibfnamefont {A.}~\bibnamefont {Kozakov}}, \bibinfo {author}
  {\bibfnamefont {A.}~\bibnamefont {Kochur}}, \bibinfo {author} {\bibfnamefont
  {A.}~\bibnamefont {Nikolskii}}, \bibinfo {author} {\bibfnamefont
  {I.}~\bibnamefont {Petrov}}, \bibinfo {author} {\bibfnamefont
  {S.}~\bibnamefont {Shevtsova}},\ and\ \bibinfo {author} {\bibfnamefont
  {N.}~\bibnamefont {Ayachit}},\ }\bibfield  {title} {\bibinfo {title} {Crystal
  structure, valence state of ions and magnetic properties of
  $\mathrm{HoFeO}_{3}$ and $\mathrm{HoFe}_{0.8}\mathrm{Sc}_{0.2}\mathrm{O}_{3}$
  nanoparticles from $\mathrm{X}$-ray diffraction, $\mathrm{X}$-ray
  photoelectron, and $\mathrm{M}${\"o}ssbauer spectroscopy data},\ }\href@noop
  {} {\bibfield  {journal} {\bibinfo  {journal} {Journal of Alloys and
  Compounds}\ }\textbf {\bibinfo {volume} {842}},\ \bibinfo {pages} {155805}
  (\bibinfo {year} {2020})}\BibitemShut {NoStop}%
\bibitem [{\citenamefont {Pi{\~n}a}\ \emph {et~al.}(2008)\citenamefont
  {Pi{\~n}a}, \citenamefont {Buentello}, \citenamefont {Arriola},\ and\
  \citenamefont {Nava}}]{bib33}%
  \BibitemOpen
  \bibfield  {author} {\bibinfo {author} {\bibfnamefont {P.}~\bibnamefont
  {Pi{\~n}a}}, \bibinfo {author} {\bibfnamefont {R.}~\bibnamefont {Buentello}},
  \bibinfo {author} {\bibfnamefont {H.}~\bibnamefont {Arriola}},\ and\ \bibinfo
  {author} {\bibfnamefont {E.}~\bibnamefont {Nava}},\ }\bibfield  {title}
  {\bibinfo {title} {M{\"o}ssbauer spectroscopy of lanthanum and holmium
  ferrites},\ }\href@noop {} {\bibfield  {journal} {\bibinfo  {journal}
  {Hyperfine Interactions}\ }\textbf {\bibinfo {volume} {185}},\ \bibinfo
  {pages} {173} (\bibinfo {year} {2008})}\BibitemShut {NoStop}%
\bibitem [{\citenamefont {Sternheimer}(1950)}]{bib35}%
  \BibitemOpen
  \bibfield  {author} {\bibinfo {author} {\bibfnamefont {R.}~\bibnamefont
  {Sternheimer}},\ }\bibfield  {title} {\bibinfo {title} {On nuclear quadrupole
  moments},\ }\href@noop {} {\bibfield  {journal} {\bibinfo  {journal}
  {Physical Review}\ }\textbf {\bibinfo {volume} {80}},\ \bibinfo {pages} {102}
  (\bibinfo {year} {1950})}\BibitemShut {NoStop}%
\bibitem [{\citenamefont {Sternheimer}(1951)}]{bib36}%
  \BibitemOpen
  \bibfield  {author} {\bibinfo {author} {\bibfnamefont {R.}~\bibnamefont
  {Sternheimer}},\ }\bibfield  {title} {\bibinfo {title} {On nuclear quadrupole
  moments},\ }\href@noop {} {\bibfield  {journal} {\bibinfo  {journal}
  {Physical Review}\ }\textbf {\bibinfo {volume} {84}},\ \bibinfo {pages} {244}
  (\bibinfo {year} {1951})}\BibitemShut {NoStop}%
\bibitem [{\citenamefont {Marathe$\mathrm{,}$}\ and\ \citenamefont
  {Trautwein}(1979)}]{bib37}%
  \BibitemOpen
  \bibfield  {author} {\bibinfo {author} {\bibfnamefont {L.~S. V.~R.}\
  \bibnamefont {Marathe$\mathrm{,}$}}\ and\ \bibinfo {author} {\bibfnamefont
  {A.}~\bibnamefont {Trautwein}},\ }\bibfield  {title} {\bibinfo {title}
  {Sternheimer shielding using various approximations},\ }\href@noop {}
  {\bibfield  {journal} {\bibinfo  {journal} {Physical Review A}\ }\textbf
  {\bibinfo {volume} {19}},\ \bibinfo {pages} {1852} (\bibinfo {year}
  {1979})}\BibitemShut {NoStop}%
\bibitem [{\citenamefont {Brinks}\ \emph {et~al.}(2001)\citenamefont {Brinks},
  \citenamefont {Rodr{\'\i}guez-Carvajal}, \citenamefont {Fjellv{\aa}g},
  \citenamefont {Kjekshus},\ and\ \citenamefont {Hauback}}]{bib38}%
  \BibitemOpen
  \bibfield  {author} {\bibinfo {author} {\bibfnamefont {H.}~\bibnamefont
  {Brinks}}, \bibinfo {author} {\bibfnamefont {J.}~\bibnamefont
  {Rodr{\'\i}guez-Carvajal}}, \bibinfo {author} {\bibfnamefont
  {H.}~\bibnamefont {Fjellv{\aa}g}}, \bibinfo {author} {\bibfnamefont
  {A.}~\bibnamefont {Kjekshus}},\ and\ \bibinfo {author} {\bibfnamefont
  {B.}~\bibnamefont {Hauback}},\ }\bibfield  {title} {\bibinfo {title} {Crystal
  and magnetic structure of orthorhombic $\mathrm{HoMnO}_{3}$},\ }\href@noop {}
  {\bibfield  {journal} {\bibinfo  {journal} {Physical Review B}\ }\textbf
  {\bibinfo {volume} {63}},\ \bibinfo {pages} {094411} (\bibinfo {year}
  {2001})}\BibitemShut {NoStop}%
\bibitem [{\citenamefont {Holmes}\ \emph {et~al.}(1971)\citenamefont {Holmes},
  \citenamefont {Van~Uitert},\ and\ \citenamefont {Hecker}}]{bib39}%
  \BibitemOpen
  \bibfield  {author} {\bibinfo {author} {\bibfnamefont {L.}~\bibnamefont
  {Holmes}}, \bibinfo {author} {\bibfnamefont {L.}~\bibnamefont {Van~Uitert}},\
  and\ \bibinfo {author} {\bibfnamefont {R.}~\bibnamefont {Hecker}},\
  }\bibfield  {title} {\bibinfo {title} {Effect of co on magnetic properties of
  $\mathrm{ErFeO}_{3}$, $\mathrm{HoFeO}_{3}$, and $\mathrm{DyFeO}_{3}$},\
  }\href@noop {} {\bibfield  {journal} {\bibinfo  {journal} {Journal of Applied
  Physics}\ }\textbf {\bibinfo {volume} {42}},\ \bibinfo {pages} {657}
  (\bibinfo {year} {1971})}\BibitemShut {NoStop}%
\bibitem [{\citenamefont {Lyubutin}\ \emph {et~al.}(2011)\citenamefont
  {Lyubutin}, \citenamefont {Naumov}, \citenamefont {Mill}, \citenamefont
  {Frolov},\ and\ \citenamefont {Demikhov}}]{bib34}%
  \BibitemOpen
  \bibfield  {author} {\bibinfo {author} {\bibfnamefont {I.}~\bibnamefont
  {Lyubutin}}, \bibinfo {author} {\bibfnamefont {P.}~\bibnamefont {Naumov}},
  \bibinfo {author} {\bibfnamefont {B.}~\bibnamefont {Mill}}, \bibinfo {author}
  {\bibfnamefont {K.}~\bibnamefont {Frolov}},\ and\ \bibinfo {author}
  {\bibfnamefont {E.}~\bibnamefont {Demikhov}},\ }\bibfield  {title} {\bibinfo
  {title} {Structural and magnetic properties of the iron-containing langasite
  family $\mathrm{A}_{3}\mathrm{MFe}_{3}\mathrm{X}_{2}\mathrm{O}_{14}$
  ($\mathrm{A= Ba, Sr; M= Sb, Nb, Ta; X= Si, Ge}$) observed by
  $\mathrm{M}${\"o}ssbauer spectroscopy},\ }\href@noop {} {\bibfield  {journal}
  {\bibinfo  {journal} {Physical Review B}\ }\textbf {\bibinfo {volume} {84}},\
  \bibinfo {pages} {214425} (\bibinfo {year} {2011})}\BibitemShut {NoStop}%
\end{thebibliography}%

\end{document}